\documentclass[draft]{agujournal2019}
\usepackage{url} 
\usepackage{lineno}
\usepackage[inline]{trackchanges} 
\usepackage{soul}
\usepackage{amsmath}
\usepackage{tabularx}
\usepackage{booktabs}
\usepackage{multirow}
\usepackage{graphicx}
\usepackage{placeins}
\usepackage{color}

\UseRawInputEncoding
\draftfalse

\journalname{Journal of Advances in Modeling Earth Systems}

\begin{document}


\title{ADAF: An Artificial Intelligence Data Assimilation Framework for Weather Forecasting}


%
%



\authors{Yanfei Xiang\affil{1,2}, Weixin Jin\affil{2}, Haiyu Dong\affil{3}, Mingliang Bai\affil{3}, Zuliang Fang\affil{3}, Pengcheng Zhao\affil{3}, Hongyu Sun\affil{4}, Kit Thambiratnam\affil{4}, Qi Zhang\affil{2}, Xiaomeng Huang\affil{1}}



\affiliation{1}{Department of Earth System Science, Ministry of Education Key Laboratory for Earth System Modeling, Institute for Global Change Studies, Tsinghua University, Beijing, China}
\affiliation{2}{Microsoft, Beijing, China}
\affiliation{3}{Microsoft, Suzhou, China}
\affiliation{4}{Microsoft, Redmond, Washington, United States}

\correspondingauthor{Xiaomeng Huang}{hxm@tsinghua.edu.cn}
\correspondingauthor{Haiyu Dong}{Haiyu.Dong@microsoft.com}


\begin{keypoints}
\item An AI-based data assimilation framework (ADAF) for real-world observations from unfixed locations and multiple sources.
\item ADAF offers a simple framework to generate km-scale analysis at low cost, making it a practical component of the weather forecasting system.
\item ADAF can be adapted to low-accuracy background and extremely sparse observations and reconstruct wind fields of tropical cyclones.
\end{keypoints}

%
%

%
%

\begin{abstract}
The forecasting skill of numerical weather prediction (NWP) models critically depends on the accurate initial conditions, also known as analysis, provided by data assimilation (DA).
Traditional DA methods often face a trade-off between computational cost and accuracy due to complex linear algebra computations and the high dimensionality of the model, especially in nonlinear systems. Moreover, processing massive data in real-time requires substantial computational resources.
To address this, we introduce an artificial intelligence-based data assimilation framework (ADAF) to generate high-quality kilometer-scale analysis. 
This study is the pioneering work using real-world observations from varied locations and multiple sources to verify the AI method's efficacy in DA, including sparse surface weather observations and satellite imagery.
We implemented ADAF for four near-surface variables in the Contiguous United States (CONUS). 
The results indicate that ADAF surpasses the High Resolution Rapid Refresh Data Assimilation System (HRRRDAS) in accuracy by 16\% to 33\% for near-surface atmospheric conditions, aligning more closely with actual observations, and can effectively reconstruct extreme events, such as tropical cyclone wind fields. Sensitivity experiments reveal that ADAF can generate high-quality analysis even with low-accuracy backgrounds and extremely sparse surface observations.  
ADAF can assimilate massive observations within a three-hour window at low computational cost, taking about two seconds on an AMD MI200 graphics processing unit (GPU). 
ADAF has been shown to be efficient and effective in real-world DA, underscoring its potential role in operational weather forecasting.
\end{abstract}

\section*{Plain Language Summary}

Numerical Weather Prediction (NWP) models require precise initial conditions, known as analysis, to obtain accurate forecasts. This is achieved through data assimilation (DA), which combines different types of observations. Traditional DA methods struggle to balance computational cost and accuracy because of the complex mathematical operations involved and the need to handle a lot of information quickly. To solve this, the Artificial Intelligence-based Data Assimilation Framework (ADAF) was created. ADAF uses a deep learning model to improve the accuracy of analysis, using observations from various sources and locations such as weather stations and satellites. 
When evaluated in the Contiguous United States (CONUS), ADAF exhibits 16\% to 33\% higher accuracy than the state-of-the-art DA system in depicting near-surface atmospheric conditions and can reconstruct atmospheric states during severe wind events such as tropical cyclones.
It performs well even with limited data. ADAF can process large amounts of data at a low computational cost. 
This study demonstrates the effectiveness of the AI method in the real-world DA scenario, showing great promise to make weather forecasts more efficient and reliable.

%
%

\section{Introduction}

The precision of weather forecasts is highly dependent on the quality of the initial conditions provided by data assimilation (DA)~\cite{bauer_quiet_2015, Griffith2000Adjoint, Hunt2005Efficient}. 
DA methods improve initial conditions by assimilating observations into numerical weather prediction (NWP) models, increasing the reliability of the forecast~\cite{asch_data_2016}. 
Traditional DA methods are classified into variational, ensemble, and hybrid methods~\cite{Bannister2017review}. Variational methods, such as three- and four-dimensional variational assimilation (3D-Var and 4D-Var), minimize a cost function to optimize model states~\cite{Janiskova_1999_4DVar, Rabier_2000_4DVar, Pierre_2007_4DVar, Courtier1998The, Rabier1998The}. Ensemble methods, such as the Ensemble Kalman Filter (EnKF) and Particle Filter, use multiple simulations for system state estimation, supporting nonlinear and non-Gaussian models~\cite{Buehner2010aIntercomparison, Buehner2010bIntercomparison, evensen2003ensemble}. Hybrid methods combine ensemble and variational approaches to maximize strengths and overcome respective limitations, improving background error covariance matrices and localization~\cite{Lee2022Hybrid}.

Despite traditional DA methods have evolved, they often face a trade-off between computational cost and accuracy due to complex linear algebra computations and the high dimensionality of the model, especially in nonlinear systems~\cite{Boudier2020Data, Khaki2018Nonparametric}. 
In addition, processing substantial amounts of observational data in real time increases resource demand~\cite{eyre2022assimilation}. This becomes especially challenging when the model's state variables differ significantly from the observations in nature or resolution.
In practice, DA methods often rely on simplifying assumptions such as linearity and Gaussianity or approximations such as tangent linear and adjoint models~\cite{Miller1999Data, McLaughlin2002An}. 
These assumptions and approximations can limit the accuracy of DA methods~\cite{Kenta2021DA}.
Moreover, the implementation and maintenance of these sophisticated algorithms require expertise and ongoing optimization, further contributing to the cost. Balancing these factors requires thoughtful algorithm design, model simplification, and leveraging high-performance computing resource.

Machine learning (ML) techniques are increasingly being utilized to improve the accuracy and efficiency of traditional DA methods~\cite{Massimo2021Machine, Penny2022Integrating, sonnewald2021bridging, howard2024machine, huang2024diffda, wang2024four}. These efforts focus on creating cost-effective ML models to simplify the formulation of traditional DA methods, thus improving computational feasibility.
Several studies have highlighted the connection between DA and ML~\cite{geer_learning_2021, abarbanel_machine_2018, bocquet_data_2019}. In the Bayesian framework, DA and ML are unified in that they both utilize prior knowledge to solve an inverse problem~\cite{sonnewald2021bridging}. 
Recent progress includes the development of inverse observation operators~\cite{frerix2021variational}, 
the estimation of error covariance matrices~\cite{cocucci_model_2021, Melinc_2024}, 
the combination of traditional DA methods with surrogate AI-based dynamic models~\cite{Li2023RapidSimulation, Ashesh2022Deep}, 
the joint training of surrogate models for the dynamic system along with the DA solver~\cite{legler_combining_2022, arcucci_deep_2021, fablet_learning_2021}, 
the derivation of tangent-linear and adjoint models~\cite{Hatfield_2021_building, Xiao2024FengWu} and
the AI models for assimilating observations ~\cite{huang2024diffda, chen2024endtoend, xu2024fuxida, keller2024AI, andrychowicz2023deep}.


Idealized scenarios, such as Lorenz attractor, quasi-geostrophic, and barotropic vorticity models, are commonly used in previous studies to test methodological effectiveness~\cite{brajard_combining_2020, Yasuda_2023_Spatial, howard2024machine}. These scenarios are overly simplified, thus misrepresenting the complexities of actual atmospheric and oceanic systems. Operating at broader scales, they miss finer details and fail to capture the unpredictable nature of real-world data, which may cause AI models to excel in experimental settings but struggle in practice.
Furthermore, most previous studies are based on pseudo-observations~\cite{Xiao2024FengWu, huang2024diffda, wang_deep_2022}, which are derived by sampling and adding noise to the reanalysis data. However, pseudo- and real-world observations differ in spatial distribution, variables, and quality. Real-world data are crucial for verifying the efficacy of the AI methods in DA.

Despite the potential to predict future states directly from observations, practical applications still depend on the initial field for a 24-hour forecast~\cite{andrychowicz2023deep}. This dependency exists because atmospheric analysis offers a more complete and accurate representation than individual observations~\cite{Houtekamer2001A}. DA can correct errors and inconsistencies found in raw observations, reducing random noise, and providing a smoother, more reliable representation. In addition, the analysis field fills spatial and temporal gaps between sparse observations, ensuring continuous coverage across regions and time.
Thus, the analysis field is indispensable for accurate weather forecasting, serving as initial conditions in both traditional numerical and AI forecasting models~\cite{pathak_fourcastnet_2022, bi_accurate_2023, lam_learning_2023, chen_fuxi_2023, chen2023fengwu}.

In this study, we introduce an AI-based data assimilation framework (ADAF) to generate high-quality kilometer-scale analysis. 
This study is the pioneering work using multisource real-world observations to verify the AI method's efficacy in DA, including sparse surface weather observations and satellite imagery. ADAF can handle surface location-varying observations effortlessly.
We implemented ADAF for four near-surface variables in the Contiguous United States (CONUS). 
The results indicate that ADAF surpasses HRRRDAS in representing near-surface atmospheric conditions, demonstrating closer alignment with real observations. ADAF shows accuracy enhancements of 16\% to 33\% over RTMA and 5.7\% to 7.7\% relative to observations.
Sensitivity experiments reveal that ADAF can generate high-quality analysis even with low-accuracy backgrounds and extremely sparse surface observations. 
ADAF can reconstruct tropical cyclone wind fields and assimilate massive observations within a three-hour window at low computational cost, taking about two seconds on an AMD MI200 GPU. 
ADAF has been proven efficient and effective in real-world DA, highlighting its potential in operational weather forecasting.

This paper is structured as follows. In Section~\ref{sec:methods}, we describe the general concept of DA and the proposed AI-based data assimilation framework (ADAF). The pipeline of ADAF and the architecture of the neural network are detailed. In Section~\ref{sec:experiments}, we present the data used in this study, the implementation of the experiment, and the evaluation approaches. We assessed performance by grid-to-grid and grid-to-station and compared these results with the state-of-the-art DA system. We also performed sensitivity experiments to verify the robustness of our method. In addition, some cases, including the tropical cyclone, were visualized to demonstrate the effectiveness of our method. Section~\ref{sec:conclusion_and_discussion} concludes the paper and describes the future direction of research.

\section{Methods}
\label{sec:methods}

\subsection{General Concept}

The Kalman Filter (KF) and its variants ~\cite{kalman1960new, Evensen_EnKF_1994} are important DA algorithms to estimate the state of a dynamic system over time.  
It consists of two iterative steps: predicting the future state and updating the estimate with new data.
The prediction step projects the current state estimate forward in time, utilizing the dynamic model that describes the state evolution.

\begin{linenomath*}
\begin{equation}
    \mathbf{X}^f(t)=\mathbf{M}_t\left[\mathbf{X}^a(t-1)\right]+\mathbf{h}(t)
\label{eq:enkf_forecast_step}
\end{equation}
\end{linenomath*}
where $\mathbf{X}^f(t)$ denotes the predicted state estimate at time $t$. $\mathbf{M}_t$ is the dynamic model that advances the state from time $t-1$ to $t$. $\mathbf{X}^a(t-1)$ represents the optimal state estimation (also known as analysis) at the previous time $t-1$. $\mathbf{h}(t)$ quantifies the model error, indicating the uncertainty in the dynamic model.

After obtaining the predicted state, the KF integrates observations to refine the state estimate during the analysis step. In this process, the Kalman gain is used to balance the uncertainties in the predicted state and observations. The relevant equations are as follows:

\begin{linenomath*}
\begin{equation}
    \mathbf{d}(t)=\mathbf{Y}(t)-\mathbf{H}_t\left[\mathbf{X}^f(t)\right]+\epsilon(t)
\label{eq:enkf_analysis_step_1}
\end{equation}
\end{linenomath*}

\begin{linenomath*}
\begin{equation}
    \mathbf{K}(t)=\mathbf{P}^f(t) \mathbf{H}_t^T\left[\mathbf{H}_t \mathbf{P}^f(t) \mathbf{H}_t^T+\mathbf{R}(t)\right]^{-1}
\label{eq:enkf_analysis_step_2}
\end{equation}
\end{linenomath*}

\begin{linenomath*}
\begin{equation}
    \mathbf{X}^d(t) = \mathbf{K}(t) \mathbf{d}(t) 
\label{eq:enkf_analysis_step_3}
\end{equation}
\end{linenomath*}

\begin{linenomath*}
\begin{equation}
    \mathbf{X}^a(t)=\mathbf{X}^f(t)+\mathbf{X}^d(t)
\label{eq:enkf_analysis_step_4}
\end{equation}
\end{linenomath*}
where $\mathbf{d}(t)$ is the innovation vector at time $t$, indicating the difference between observations and the predicted state. 
$\mathbf{Y}(t)$ is observations.
$\mathbf{H}_t$ is the observation operator that maps the model's state variables to the observation space.
$\mathbf{K}(t)$ is the Kalman gain. $\mathbf{P}^f(t)$ is the forecast error covariance matrix indicating uncertainties in the system's predicted state.
$\mathbf{R}(t)$ is the observation error covariance matrix representing the uncertainties in observations.
$\epsilon(t)$ represents the observation noise.

The product $\mathbf{X}^d(t) = \mathbf{K}(t) \mathbf{d}(t)$ computes the adjustment that is used to refine the forecast state $\mathbf{X}^f(t)$. The optimal state $\mathbf{X}^a(t)$, known as the analysis, is obtained by adding the product to the predicted state to better align with the new observation $\mathbf{Y}(t)$.
Despite this, computing this product poses certain difficulties. A key issue is the precise determination of the Kalman gain $\mathbf{K}(t)$, which depends on accurately estimating the forecast and observation error covariance matrices ($\mathbf{P}^f$ and $\mathbf{R}$). This is challenging because of their dynamic properties and the computational demands of matrix operations, particularly in high-dimensional contexts~\cite{Furrer2007Estimation, Zhen2014Adaptive, Liang2012Maximum, Furrer2007Estimation}. 
Additionally, computing the innovation vector $\mathbf{d}(t)$ is crucial and can be hindered by nonlinear observation operators and the quality and sparsity of observations~\cite{tandeo_review_2020, carrio2019potential}.

To overcome the computational challenges in traditional EnKF methods, our research aims to use a neural network $\mathcal{F}_{NN}$ to approximate the product $\mathbf{K}(t) \mathbf{d}(t)$, offering a data-driven substitute for conventional DA methods, which can be formulated as follows:

\begin{linenomath*}
\begin{equation}
    \mathcal{F}_{NN}(\mathbf{X}^f(t), \mathbf{Y}(t); \theta) = \mathbf{X}^d(t)
\label{eq:adjustment_learning_1}
\end{equation}
\end{linenomath*}

\begin{linenomath*}
\begin{equation}
    \mathbf{X}^a(t)=\mathbf{X}^f(t)+ \mathcal{F}_{NN}(\mathbf{X}^f(t), \mathbf{Y}(t); \theta)
\label{eq:adjustment_learning_2}
\end{equation}
\end{linenomath*}
where $\theta$ denotes the model parameters of $ \mathcal{F}_{NN}$.

\subsection{ADAF: AI-based Data Assimilation Framework}
\label{sec:Network}

\begin{figure}[h!]
    \centering    
    \includegraphics[width=1\textwidth]{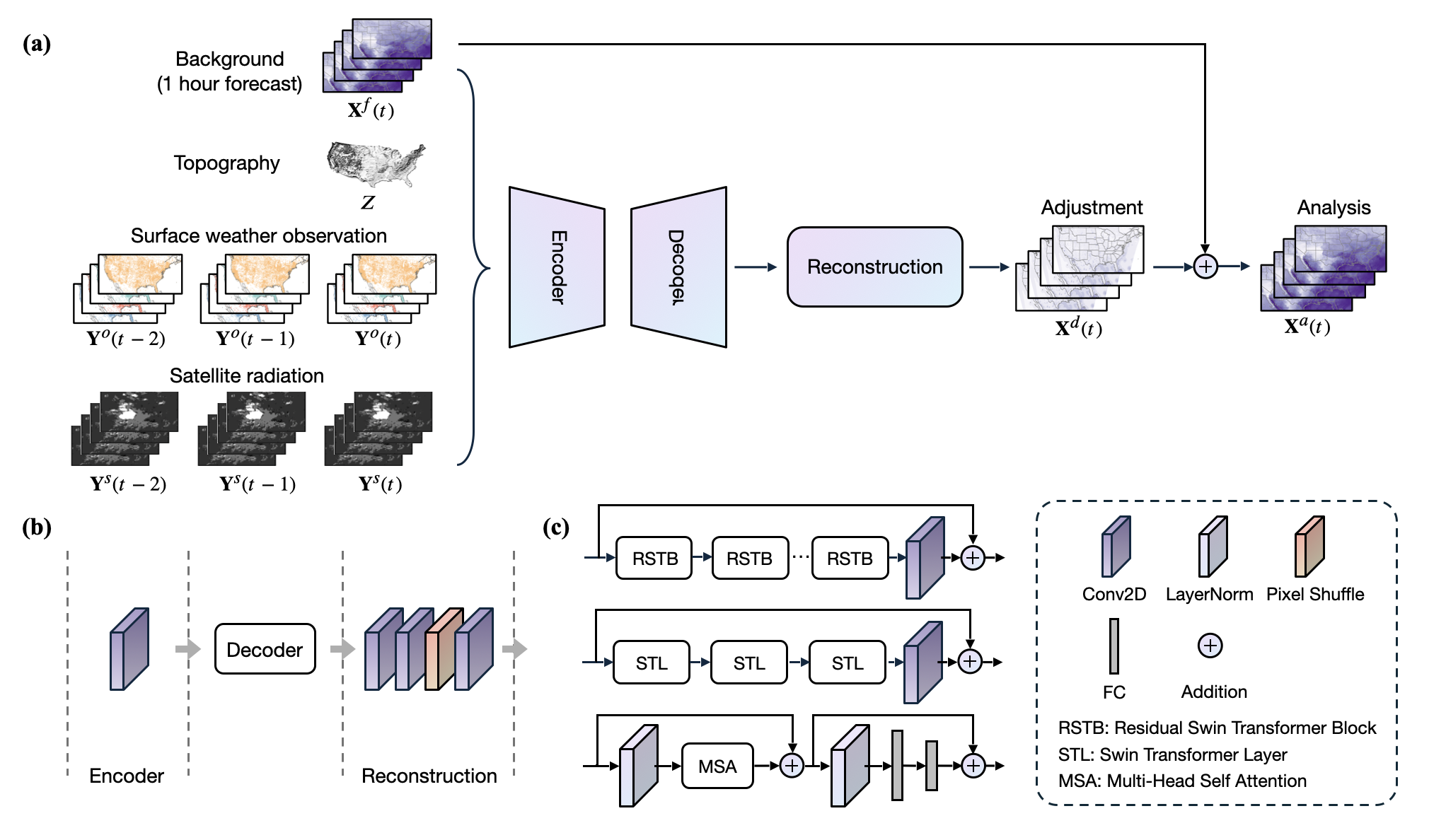}
    \caption{An overview of ADAF and the architecture of neural networks used in ADAF. (a) An overview of ADAF. The input consists of three types of data, including observations within 3-hour window, background, and topography. The output is the analysis. (b) The neural network architecture implemented in ADAF consists of an encoder, a decoder, and a reconstruction module. (c) Details of the decoder, which includes residual Swin Transformer blocks (RSTB), Swin Transformer layers (STL), and Multi-head Self Attention (MSA).}
    \label{fig:Network_Arch}
\end{figure}

Figure~\ref{fig:Network_Arch}(a) illustrates an overview of ADAF. The inputs encompass three types of data: observations, background, and topography. Observations include surface weather observations $\mathbf{Y}^o$ and satellite imagery $\mathbf{Y}^s$ within the $T$-hour window. The background $\mathbf{X}^f(t)$ derives from the 1-hour forecast at time $t$. The output of the neural network is the analysis $\mathbf{X}^a(t)$ at time $t$. 
Various multimodal data, including specific humidity, topography, and radiation, often serve as indirect proxies for target states like air temperature. Capturing these relationships is challenging due to their varying spatial coverage and dimensions. Although a mapping function, such as the observation operator that bridges heterogeneous state-observation spaces, is available, imperfect knowledge can amplify errors in high-dimensional states, necessitating further calibrations to correct biases~\cite{Jianyu2023Amachine, qu2024deep}.

Deep learning overcomes this challenge by learning a function that maps both states and observations into a unified, reduced-order latent space. In this study, we used an encoder that inputs a concatenation of background, topography, and each observation modality along the channels. The decoder then maps the latent features extracted via the encoder to the state space.
Subsequently, the reconstruction module reconstructs the adjustment $\mathbf{X}^d(t)$. Finally, this adjustment is added to the background $\mathbf{X}^f(t)$, resulting in the analysis $\mathbf{X}^a(t)$.

First, we used an encoder to extract latent features from the inputs. This procedure is depicted in Equation~\ref{eq:encoder}, which uses a 2D convolution layer (Conv2D) with a kernel size of $3 \times 3$. 

\begin{linenomath*}
\begin{equation}
    F_0 = H_{\text{enc}}(\mathbf{Y}^o, \mathbf{Y}^s, \mathbf{X}^f(t), Z)
\label{eq:encoder}
\end{equation}
\end{linenomath*}
Where $H_{\text{enc}}$ is the function of encoder. $F_0$ are the latent features. $\mathbf{Y}^o = \{ \mathbf{Y}^o(t-i) \}, i=0,1,...,T$ and $\mathbf{Y}^s = \{ \mathbf{Y}^s(t-i) \}, i=0,1,...,T$ are the surface weather observations and satellite radiation within $T$-hour window.
$Z$ denotes topography. $\mathbf{X}^f(t)$ represents the background. 

The latent features extracted by the encoder are fed into the decoder $H_{\text{dec}}(\cdot)$, as depicted in the following equations:
\begin{linenomath*}
\begin{equation}
    F_{\text{DF}} = H_{\text{dec}}(F_0)
\label{eq:decoder_2}
\end{equation}
\end{linenomath*}
where $F_0$ denotes the combined features. $F_{\text{DF}}$ denotes the deep features decoded by the decoder $H_{\text{dec}}(\cdot)$.
The decoder uses residual Swin transformer blocks (RSTB)~\cite{liang2021swinir} and a 2D convolution layer to learn from the fusion features of the observations and the background, to capture high-frequency information, as illustrated in Figure~\ref{fig:Network_Arch}(c).
Employing a long skip connection allows the model to transmit low-frequency information directly to the reconstruction module. This enables the module to focus on high-frequency data, thus stabilizing the training process.
The intermediate features $F_1, F_2, F_3$ and the final deep feature $F_{\text{DF}}$ are decoded sequentially as follows:

\begin{linenomath*}
\begin{equation}
    F_{i} = H_{\text{RSTB}_i}(F_{i-1}), i = 1,2,3
\label{eq:decoder_detail_1}
\end{equation}
\end{linenomath*}

\begin{linenomath*}
\begin{equation}
    F_{\text{DF}} = H_{\text{Conv}}(F_3) + F_0
\label{eq:decoder_detail_2}
\end{equation}
\end{linenomath*}
where $H_{RSTB_i}(\cdot)$ indicates the $i^{th}$ RSTB and $H_{Conv}(\cdot)$ is the final convolutional layer. $F_{\text{DF}}$ is a feature map with 64 channels, which is forwarded to the reconstruction module. 
The RSTB comprises Swin Transformer layers (STL) and convolutional layers, as illustrated in Figure~\ref{fig:Network_Arch}(c). 
The residual connection offers an identity-based linkage between various blocks and the reconstruction module, facilitating the aggregation of features at different levels. 
STL \cite{Liu_2021_Swin_transformer} includes multi-head self-attention (MSA) \cite{Vaswani_2017_Attention}, layer normalization (LayerNorm), fully connected layers (FC), and a residual connection, as shown in Figure~\ref{fig:Network_Arch}(c).

The reconstruction module is used to reconstruct the adjustment of states by combining features of the encoder and decoder, as described in Equation~\ref{eq:reconstruct_module}. 
This module includes pixel shuffle layers \cite{Shi_sub_pixel_2016} and convolution layers with a kernel size of $3 \times 3$, as illustrated in Figure~\ref{fig:Network_Arch}(b). 

\begin{linenomath*}
\begin{equation}
    \mathbf{X}^d(t) = H_{rec}(F_0 + F_{\text{DF}})
\label{eq:reconstruct_module}
\end{equation}
\end{linenomath*}
where $H_{rec}(\cdot)$ is the function of the reconstruction module. $\mathbf{X}^d(t)$ is the adjustment. Finally, the analysis is produced by adding adjustment and background using the following equation:

\begin{equation}
\begin{aligned}
   \mathbf{X}^a(t) = \mathbf{X}^d(t) + \mathbf{X}^f(t)
\end{aligned}
\label{eq:adjustment_add}
\end{equation}

\section{Experiments}
\label{sec:experiments}

\subsection{Data Preparation}
\label{sec:data_preparation}

\subsubsection{Types of Datasets}
\label{sec:types_of_datasets}

The data used in this study can be grouped into four categories: forecast, analysis, observations, and static topography data. 
Forecast data is generated by High Resolution Rapid Refresh (HRRR)~\cite{Dowell_2022_HRRR}, a 3 km resolution NWP model developed by the National Oceanic and Atmospheric Administration (NOAA). 
HRRRv4 is the latest version, released in December 2020, exhibits a significant improvement in DA using the HRRR 3-km data assimilation system (HRRRDAS). The geopotential height from the European Centre for Medium-Range Weather Forecasts Reanalysis v5 (ERA5)~\cite{hersbach2019era5} is used as the topography.

The analysis generated by HRRRDAS is used as initial conditions in HRRRv4. In HRRRDAS, the DA process uses the Ensemble Kalman Filter (EnKF) to assimilate traditional observations, satellite data, and radar reflectivity. More details can be found in~\ref{SI-sec:HRRRDAS}.
Another set of analysis data is sourced from the Real-Time Mesoscale Analysis (RTMA)~\cite{Maunel_2011_RTMA}, which offers a high spatial and temporal resolution analysis of near-surface weather conditions. RTMA provides hourly analysis with a resolution of 2.5 km in the CONUS.

We used two types of observations: surface weather observations and satellite imagery. 
Surface weather observations are sourced from WeatherReal-Synoptic~\cite{jin2024weatherreal}, which gives quality control to station measurements collected by Synoptic Data.
For more details, please visit the Synoptic Data official website~\cite{synoptic}.
The average number of observations at each hour is illustrated in Figure~\ref{SI-fig:number_of_meansurements}. And the spatial distribution is shown in Figure~\ref{SI-fig:spatial_dist_of_meansurements}.
Satellite imagery is obtained from Geostationary Operational Environmental Satellite-16 (GOES-16)~\cite{tan2019goes}.
The Advanced Baseline Imager (ABI) on the satellite provides high spatial and temporal resolution in multiple spectral bands, comprising wavelengths from visible to infrared with a spatial resolution of approximately 1 km and a temporal resolution of approximately 15 minutes. This study utilized 2, 7, 10, and 14 bands with central shortwaves of 0.64, 3.9, 7.3, and 11.2 $\mu m$, associated with winds, clouds, water vapor, and rainfall.


\subsubsection{Data Processing}

Our research focuses on the CONUS, which is bounded by $24.70 ^\circ $ to $50.25 ^\circ N$, $64.00 ^\circ$ to $128.00 ^\circ W$. 
Near-surface meteorological variables are crucial for understanding and predicting various environmental and climatic processes that directly impact human activities.
We focused on four near-surface meteorological variables, including 2 meter temperature (T2M), specific humidity (Q), 10 meter u-component of wind (U10) and 10 meter v-component of wind (V10). 
All data were regularized to grids of size 512$\times$1280 with a spatial resolution of 0.05 $\times$ 0.05 $^\circ$. 
The surface weather observations were mapped to the nearest grids. 
The training dataset of the AI model consists of input-target pairs. Table~\ref{tab:datasets_summary} provides a summary of the input and target datasets used in our study. The inputs include surface weather observations within a 3-hour window, GOES-16 satellite imagery within a 3-hour window, background from HRRR forecast, and topography. The target combines RTMA with surface weather observations, using observations at measured locations and RTMA for locations without observations.
We used the difference between the background and target as the label during the training phase.

\renewcommand{\arraystretch}{1.5} 
\begin{table}[h!]
\caption{Overview of the datasets used for input and target during the training of the AI model. All data were regularized to grids of size 512 $\times$ 1280 with a spatial resolution of $0.05 \times 0.05 ^\circ$. }
\centering
\label{tab:datasets_summary}
\begin{tabularx}{1\textwidth}{>{\raggedright}p{1cm} >{\raggedright}p{5cm} >{\raggedright}p{2.5cm} >{\raggedright\arraybackslash}X}
\toprule
 & \textbf{Dataset} & \textbf{Time window} & \textbf{Variables/Bands} \\ 
\midrule
\multirow{4}{*}{Input} & Surface weather observations & 3 hours & Q, T2M, U10, V10 \\ 
 & GOES-16 satellite imagery & 3 hours & 0.64, 3.9, 7.3, 11.2 $\mu m$ \\ 
 & HRRR forecast & N/A & Q, T2M, U10, V10 \\ 
 & Topography & N/A & Geopotential \\ 
\midrule
\multirow{2}{*}{Target} & RTMA & N/A & Q, T2M, U10, V10 \\ 
 & Surface weather observations & N/A & Q, T2M, U10, V10 \\
\bottomrule
\end{tabularx}
\end{table}

\subsection{Implementation}

The data used in this study cover the period from October 2020 to September 2023. The data sets for training, validation and testing were split without any temporal overlap and ordered as follows: training data cover October 2020 to September 2021, validation data cover October 2021 to September 2022, and test data cover October 2022 to September 2023. 
To normalize the input and target, the max-min normalization was used. 
Given the noise in real-world data, the L1 loss remains relatively unaffected by minor data variations, resulting in more stable model predictions~\cite{Zhang2020Robust, Barron2017A}. This stability is crucial in real-world applications. Therefore, we used the L1 loss function to minimize the difference between ADAF's output and the target, as defined below:

\begin{linenomath}
\begin{equation}
    L1 =\frac{1}{C \times H \times W} \sum_{c=1}^C \sum_{i=1}^{H} \sum_{j=1}^{W} \left|\hat{\mathbf X}^d_{c, i, j}- \mathbf X^d_{c, i, j}\right|
\label{eq:loss_function}
\end{equation}
\end{linenomath}
where $\mathbf{X}^d$ represents the target adjustment and $\hat{\mathbf{X}}^d$ denotes the ADAF's output. $C$, $H$, and $W$ represent the number of variables and the number of points on the latitude and longitude grids, respectively. 

The model was trained on four AMD MI200 GPUs for a total of 750 epochs, with batch size = 4 per GPU. The entire model training process lasted approximately 36 hours. 
The AdamW optimizer~\cite{loshchilov2019decoupled} was used with parameters $\beta_1 = 0.9, \beta_2 = 0.999$ and a weight decay coefficient of $1 \times 10^{-5}$. 
We implement the ReduceLROnPlateau scheduling method \cite{pytorch_reducelronplateau} to dynamically adjust the learning rate. This approach reduces the learning rate when the chosen evaluation metric does not show improvement after a set period of patience.
After training, ADAF can perform data assimilation on a single AMD MI200 GPU in approximately 2 seconds.

\subsection{Evaluation}
\label{sec:evaluation}

To evaluate the performance of our proposed ADAF, we generate analysis at 00:00, 06:00, 12:00, and 18:00 UTC in the test dataset. 
Our study uses two evaluation approaches: grid-to-grid and grid-to-station evaluation. The grid-to-grid evaluation involves a comparison between ADAF analysis and RTMA's gridded analysis. For grid-to-station evaluation, ADAF analysis is compared with surface weather observations.
Gridded analysis data and sparse surface weather observations play different roles and originate from different methods. Analysis data is utilized to initialize models and forecasts, whereas surface weather observations offer localized atmospheric conditions in real-time. Analysis data relies heavily on sophisticated mathematical modeling and DA methods to create a synthesized view of the atmosphere, while surface weather observations rely mainly based on direct measurements and quality control. 
In the analysis, the grid value is the average value of each grid, whereas surface weather observations mean measurements specific to a local location.
Therefore, analysis data and surface observations can serve as ground truth from different perspectives. Analysis data provide a synthesized and model-based view of the atmosphere, and surface weather observations offer direct and localized measurements.

\renewcommand{\arraystretch}{1.5} 
\begin{table}[!h]
    \caption{Basic characteristic of data used in evaluation, including analysis produced by HRRRDAS, RTMA and surface weather observations from Synoptic.}
    \centering
    \begin{tabularx}{0.9\textwidth}{>{\raggedright}p{4.5cm} >{\raggedright}p{2cm} >{\raggedright}p{2cm} >{\raggedright\arraybackslash}X}
        \toprule
         \textbf{Data type} & \textbf{Source} & \textbf{Spatial Resolution} & \textbf{Temporal Resolution} \\
        \toprule
         Analysis & HRRRDAS & 0.05$^\circ$ & Hourly  \\
         Analysis & RTMA & 0.05$^\circ$ & Hourly  \\
        Surface weather observations & Synoptic & N/A & Hourly  \\
        \bottomrule
    \end{tabularx}
\label{tab:basline_and_ground_truth}
\end{table}

Table~\ref{tab:basline_and_ground_truth} outlines the main characteristics of the datasets used in the evaluation. For the grid-to-station evaluation, we employ cross-validation techniques~\cite{Maunel_2011_RTMA}. Several surface weather observations are withheld during the analysis generation process. Then we used the withheld observations to evaluate the accuracy of the ADAF analysis. Performance was evaluated using the following metrics: The root mean square error (RMSE), the mean absolute error (MAE), and the correlation coefficient (CORR). The equations for these metrics are as follows:

\begin{linenomath}
\begin{equation}
    RMSE = \sqrt{\frac{1}{N} \sum_{i=1}^N\left(G_i - E_i \right)^2}
\label{eq:metrics_RMSE}
\end{equation}
\end{linenomath}

\begin{linenomath}
\begin{equation}
    MAE = \frac{1}{N} \sum_{i=1}^N | G_i - E_i |
\label{eq:metrics_MAE}
\end{equation}
\end{linenomath}

\begin{linenomath}
\begin{equation}
    CORR = \frac{\sum_{i=1}^N\left(G_i - \Bar{G} \right)\left(E_i - \Bar{E} \right)}{\sqrt{\sum_{i=1}^N\left(G_i - \Bar{G} \right)^2} \sqrt{\sum_{i=1}^N\left(E_i - \Bar{E} \right)^2}}
\label{eq:metrics_CORR}
\end{equation}
\end{linenomath}
where $G$ denotes the actual values and $E$ represents the estimates values. $N$ is the number of samples. $\Bar{E}$ and $\Bar{G}$ are the average values of $E$ and $G$, respectively.

\subsection{Results}

\subsubsection{Accuracy of ADAF analysis}

To demonstrate the advantages of ADAF over traditional DA methods, we evaluated the accuracy of the analysis generated by ADAF and HRRRDAS. 
HRRRDAS is the state-of-art DA system used in HRRRv4. It uses the ensemble Kalman filter (EnKF)~\cite{Houtekamer_Review_2016}, specifically the ensemble square root filter (EnSRF)~\cite{EnsembleSquareRootFilters}, to integrate a 1-hour HRRR forecast and observations. HRRRDAS utilizes observations from different sources, such as surface measurements, GOES-16 satellite imagery, and radar reflectivity data from NOAA's Multi-Radar Multi-Sensor (MRMS) project. Further information is available in Section~\ref{SI-sec:HRRRDAS}.
Similarly to HRRRDAS, we performed DA using the proposed ADAF by combining a 1-hour HRRR forecast and observations, including surface weather observations from Synoptic and four channels of satellite imagery from GOES-16.
The main difference between HRRRDAS and ADAF is their DA algorithms and observations. HRRRDAS uses EnKF, while ADAF uses a neural network (see Section~\ref{sec:Network}). And HRRRDAS assimilate three types of observation: surface measurements, 16 channels of GOES-16 satellite imagery, and radar reflectivity data. ADAF only uses two types of observation: four channels of GOES-16 satellite imagery and surface weather observations. 
Further information about the data is provided in Section~\ref{sec:data_preparation}. It is important to note that more observation types are allowed to be assimilated in ADAF. Our study only validates the feasibility of the AI method to assimilate multisource observations.
We apply ADAF to generate the analysis for four near-surface variables: T2M, Q, U10 and V10 at 00:00, 06:00, 12:00, and 18:00 UTC. A year-long data assimilation was performed on the test dataset, covering October 2022 to September 2023. The domain-averaged metrics were computed using two evaluation approaches mentioned in Section~\ref{sec:evaluation}. 

\begin{figure}[h!]
    \centering
    \includegraphics[width=1\textwidth]{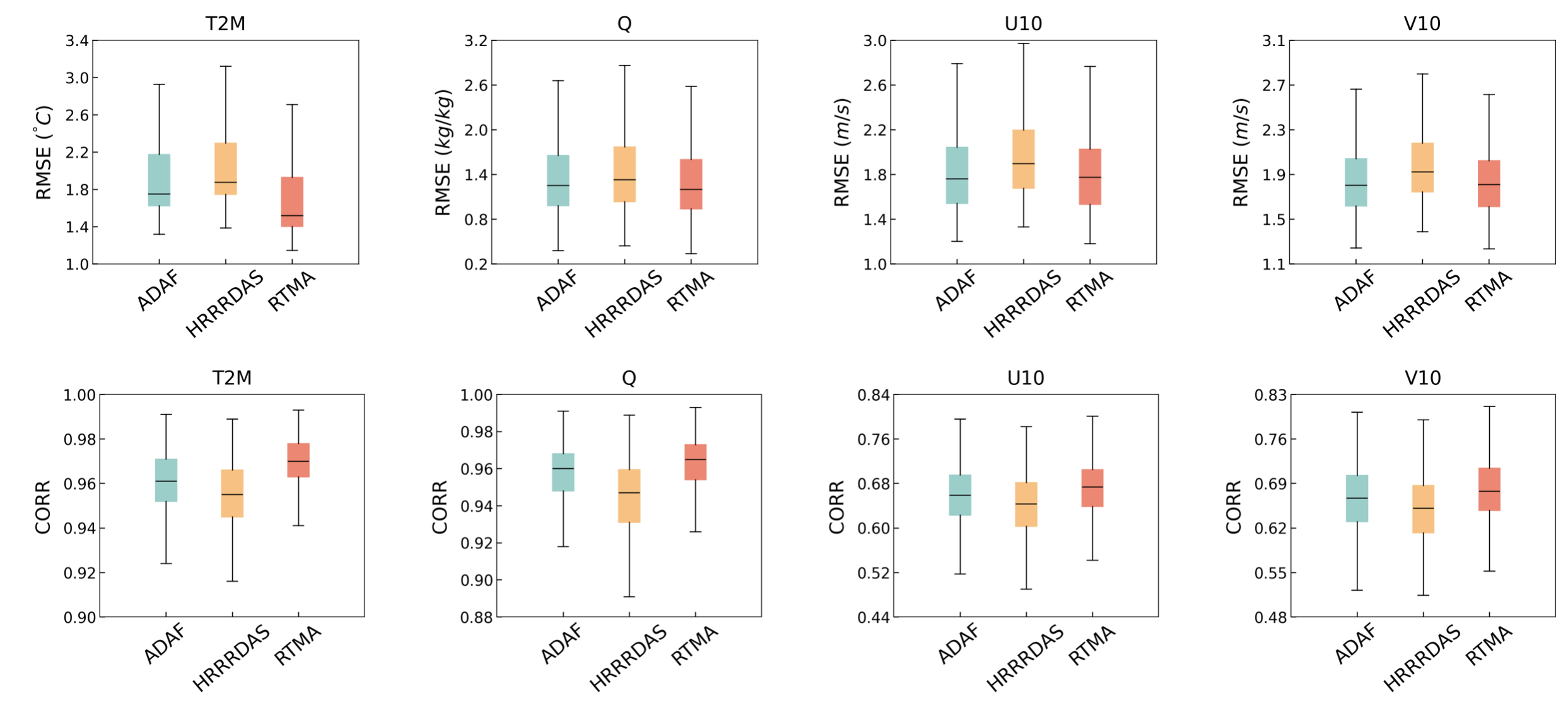}
    \caption{Domain-averaged root mean squared errors (RMSE) and correlation (CORR) for ADAF analysis (denoted as ADAF), HRRRDAS analysis (denoted as HRRRDAS) and RTMA for four near-surface variables: T2M, Q, U10 and V10. The evaluation is performed by comparing with withheld surface weather observations. The results indicate that ADAF analysis have better accuracy with lower RMSE and higher CORR than HRRRDAS analysis. ADAF analysis aligns more closely with observations than HRRRDAS. The mean RMSE values of HRRRDAS analysis for T2M, Q, U10, and V10 are 1.81, 1.21, 1.96, and 1.98. The ADAF analysis shows lower RMSE value: 1.67, 1.14, 1.81, and 1.83, showing improvements of 7.7\%, 5.7\%, 7.7\%, and 7.6\%.}
    \label{fig:overall_metrics_against_hold_obs}
\end{figure}

\begin{figure}[h!]
    \centering
    \includegraphics[width=1\textwidth]{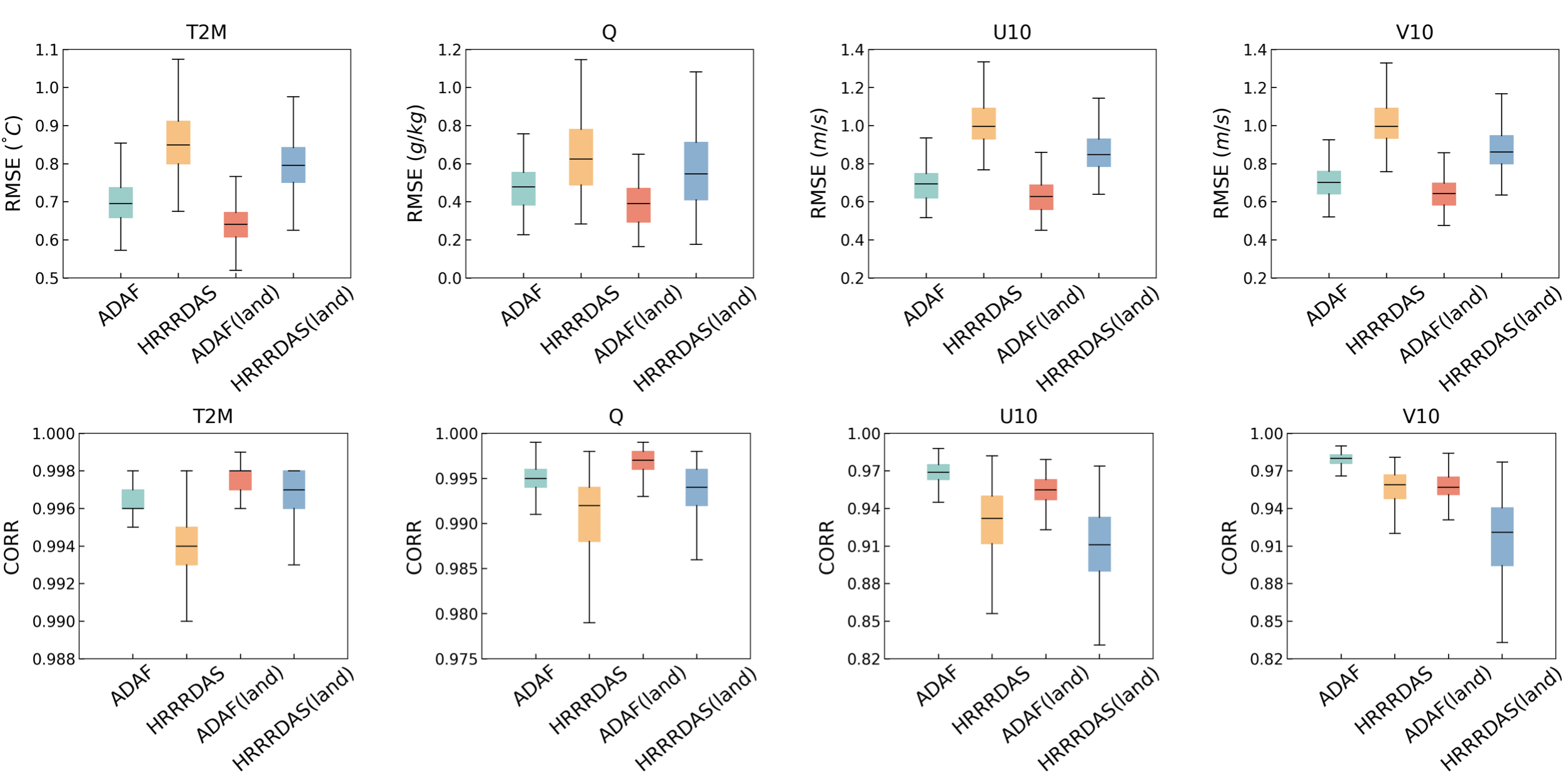}
    \caption{Domain-averaged Root Mean Squared Errors (RMSE) and Correlation (CORR) for ADAF analysis (denoted as ADAF) and HRRRDAS analysis (denoted as HRRRDAS) for four near-surface variables: T2M, Q, U10, and V10. The evaluation is performed by comparing with RTMA. 'Land' indicates that metrics are computed only over land.
    The results indicate that ADAF analysis surpasses HRRRDAS analysis in depicting surface atmospheric conditions.    
    The decrease in errors over land compared to the full study area is attributed to ADAF's efficient assimilation of surface  observations. Additionally, the error variability in ADAF analysis is less than that in HRRRDAS analysis, underscoring the robustness of ADAF.}
    \label{fig:overall_metrics_against_rtma}
\end{figure}

In grid-to-station validation, we compared ADAF analysis, HRRRDAS analysis, and RTMA with surface weather observations using cross-validation techniques.
In every DA process, approximately 10\% of surface observations (around 1600 stations) were randomly withheld from ADAF's input. 
Figure~\ref{fig:overall_metrics_against_hold_obs} illustrates the errors in ADAF analysis, HRRRDAS analysis, and RTMA. 
The ADAF analysis shows a lower mean RMSE compared to the HRRRDAS analysis: 1.67, 1.14, 1.81, and 1.83 for T2M, Q, U10, and V10, respectively, versus 1.81, 1.21, 1.96, and 1.98 from HRRRDAS, with improvements of 7.7\%, 5.7\%, 7.7\%, and 7.6\%.
The ADAF analysis demonstrates a higher CORR compared to HRRRDAS, indicating a closer alignment with observations.
The CORR values for U10 and V10 are lower than those for T2M and Q, highlighting the challenge of reconstructing the wind state in DA accurately. This challenge stems from the inherent complexity of atmospheric dynamics, which is due to nonlinear interactions among different atmospheric components.
Overall, the results demonstrate that the ADAF analysis is more closely aligned with actual observations than the HRRRDAS analysis, indicating that ADAF can effectively extrapolate sparse observations to provide reliable data for unobserved areas.
Furthermore, ADAF can process surface observations with varied locations without effort. This is because the observations' locations in the input are not fixed during the training and testing phases. This feature increases ADAF's suitability for real-world conditions, where measurement's location often vary.

The errors in the ADAF analysis are higher than those in RTMA, possibly because RTMA utilized all observations, whereas ADAF used only a subset. 
RTMA employs the two-dimensional variational analysis method (2D-Var)~\cite{Maunel_2011_RTMA}, which requires careful tuning of error covariance parameters for accurate results, particularly with low-quality background fields~\cite{Vogelzang2009Validation}. However, ADAF can be easily adapted to low-accuracy backgrounds without careful adjustments, as shown in Section~\ref{sec:sensitive_to_background_acc}.
In 2D-Var, computational cost is significantly influenced by both model resolution and data volume, with higher resolution enhancing DA performance but leading to a cubic increase in computational cost~\cite{barthelemy2022super, Yasuda_2023_Spatial}. On the other hand, ADAF remains mostly unaffected by the volume of data, as the data are pre-processed into a tensor for efficient GPU processing~\cite{Zhang2016Deep}.

\begin{figure}[htb]
    \centering
    \includegraphics[width=1\textwidth]{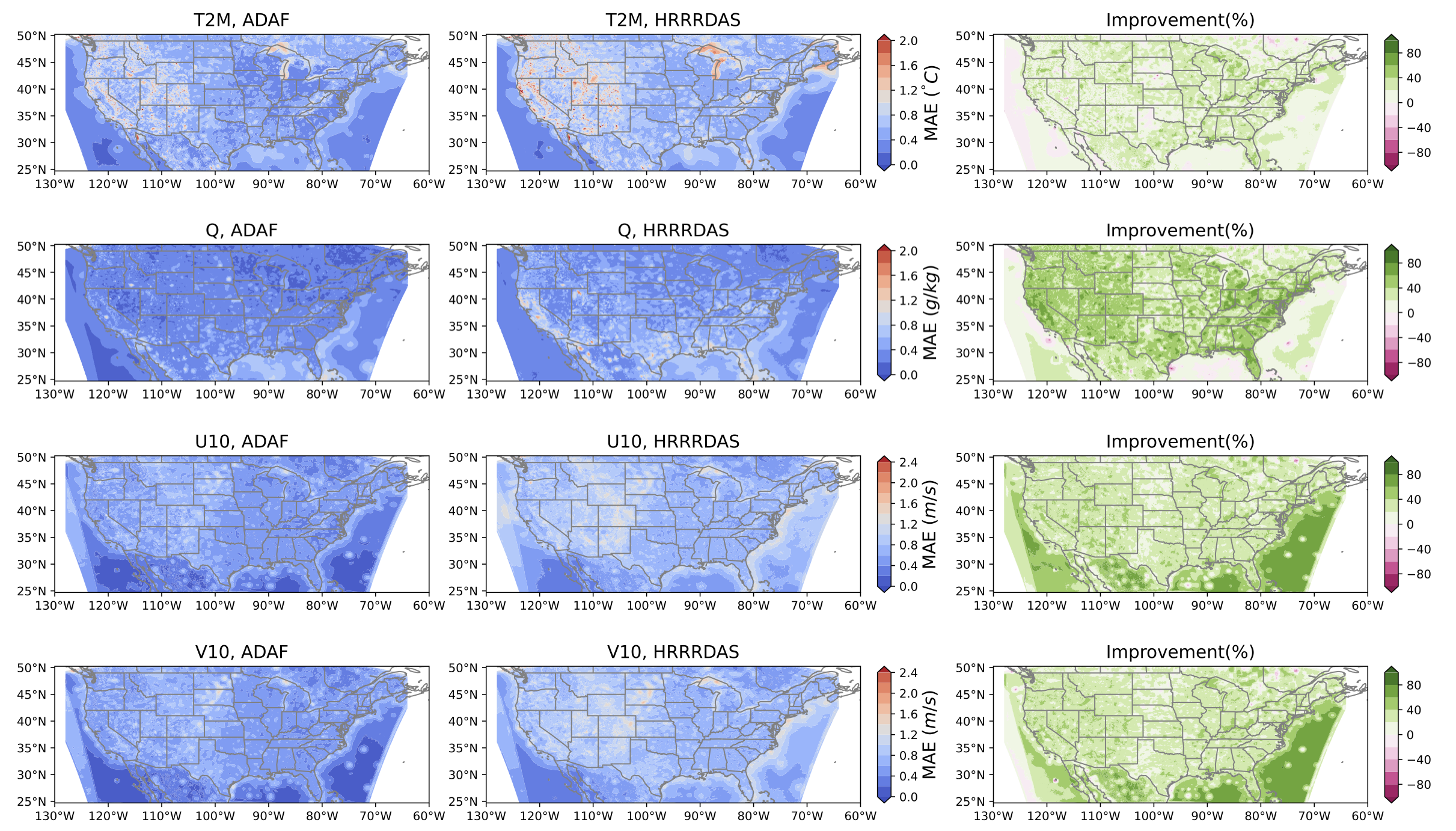}
    \caption{The Mean Absolute Errors (MAE) for ADAF and HRRRDAS analysis, compared with RTMA. 
    The first and second columns present the MAE spatial distribution for ADAF and HRRRDAS analysis, respectively. The third column illustrates the ratio of MAE reduction, calculated as $(MAE_{\text{HRRRDAS}} - MAE_{\text{ADAF}}) / MAE_{\text{HRRRDAS}}$. In most areas, ADAF analysis generally shows lower MAE compared to HRRRDAS analysis. About 82\%, 86\%, 93\%, and 92\% of the areas show improvements, with average MAE reduction of 16\%, 26\%, 35\%, and 33\% for T2M, Q, U10, and V10, respectively.}
    \label{fig:MAE_spatial_dist}
\end{figure}

In grid-to-grid validation, we evaluated the errors of the ADAF analysis and HRRRDAS analysis compared to RTMA, as shown in Figure~\ref{fig:overall_metrics_against_rtma}.
The ADAF analysis shows reduced errors compared to the HRRRDAS analysis, presenting mean RMSE values of 0.70, 0.48, 0.69, and 0.70 for T2M, Q, U10, and V10, respectively, versus 0.86, 0.64, 1.04, and 1.03 for HRRRDAS, reflecting improvements of 18.6\%, 25.0\%, 33.7\%, and 32.0\%.
In particular, significant improvements are observed on land, with mean RMSE values of 0.64, 0.39, 0.63, and 0.64 from the ADAF analysis compared to 0.80, 0.57, 0.89, and 0.91 from the HRRRDAS analysis for T2M, Q, U10, and V10, respectively. 
The percentage of improvements is 20.0\% for T2M, 31.6\% for Q, 29.2\% for U10, and 29.7\% for V10. 
The reduced errors over land versus the entire region can be attributed to ADAF's effective assimilation of surface observations on land.
In ADAF analysis, the mean CORR values are 1.0, 1.0, 0.97, and 0.98 for T2M, Q, U10, and V10, respectively. For the HRRRDAS analysis, the values are 0.99, 0.99, 0.93, and 0.96. This indicates that the ADAF analysis generally shows a higher CORR, which implies better alignment with RTMA. 
Beyond the mean RMSE and CORR values, it is also important to note that the error variability for the ADAF analysis is smaller than for the HRRRDAS analysis, indicating the robustness of ADAF.

In addition, we evaluate the spatial distribution of errors in the analysis produced by ADAF and HRRRDAS. Figure~\ref{fig:MAE_spatial_dist} shows the spatial distribution of the mean absolute error (MAE) compared to RTMA for four near-surface variables: T2M, Q, U10 and V10.
Our findings indicate that ADAF analysis generally shows lower errors compared to HRRRDAS analysis in most areas. 
About 82\%, 86\%, 93\% and 92\% of the region show improvements with an average reduction in MAE of 16\%, 26\%, 35\% and 33\% for T2M, Q, U10 and V10.
Although ADAF uses fewer observation types than HRRRDAS, it provides significant advantages in describing near-surface atmospheric conditions, mainly due to its efficient assimilation of surface weather observations.
Moreover, ADAF requires minimal computational resources. After training, ADAF can perform data assimilation on a single AMD MI200 GPU in about 2 seconds.

\subsubsection{Sensitive to Observation Sparsity}
\label{sec:sensitive_to_obs_sparsity}

\begin{figure}[h!]
    \centering
    \includegraphics[width=1\textwidth]{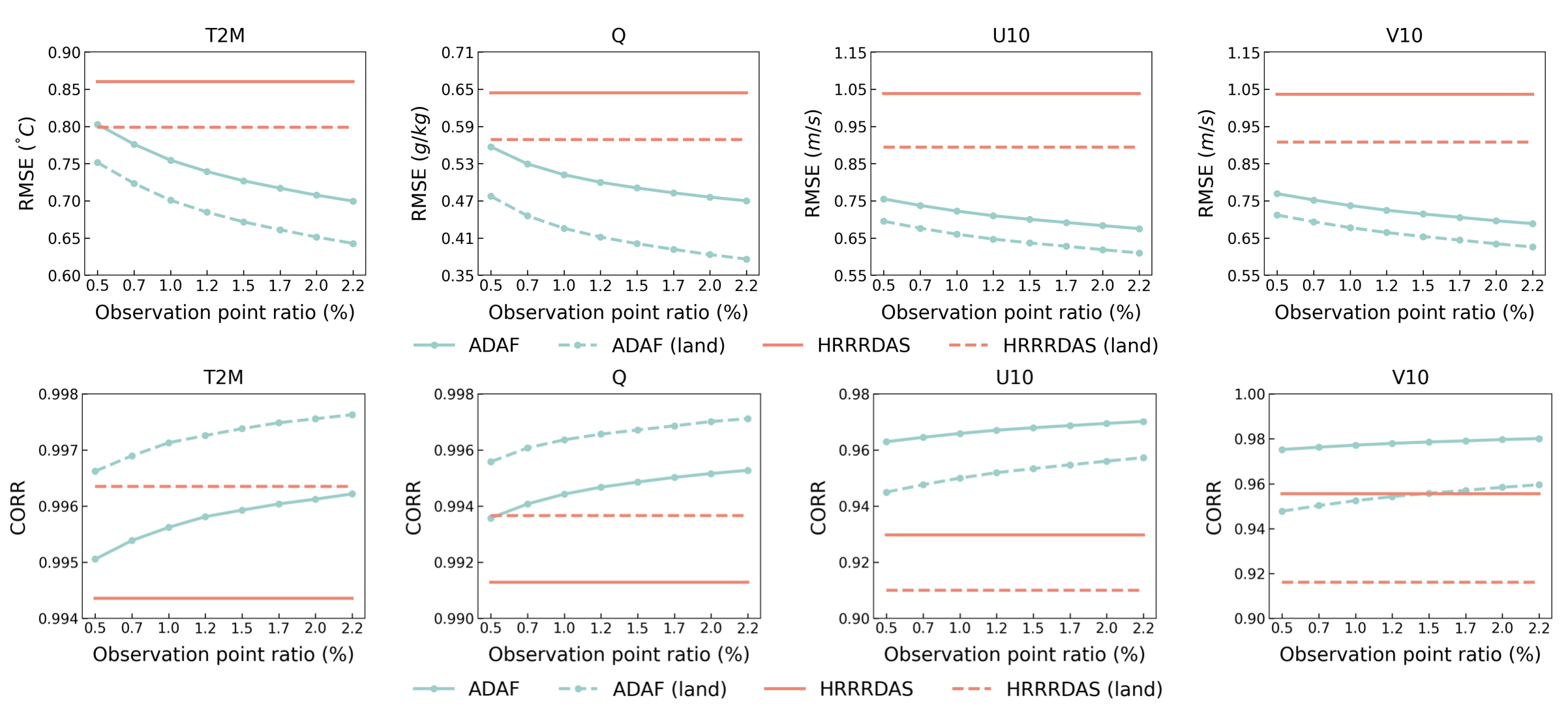}
    \caption{Relationship between analysis errors and the sparsity of observations, which is defined by the observation grid point (see Equation~\ref{eq:obs_grid_ratio}). The errors, represented as the domain-averaged RMSE and CORR, in comparison with RTMA. The ADAF analysis (depicted with blue lines) shows lower RMSE and higher CORR compared to the HRRRDAS analysis (depicted with red lines) at various observation sparsity levels. With the reduction in surface observations, the error grows, implying that a greater number of observations can improve the accuracy of the analysis. 
    The results demonstrate that ADAF is robust to generating high-quality analysis even when observations are extremely sparse (with an observation grid ratio of 0.5\%).}
    \label{fig:sensitive_to_obs_sparsity}
\end{figure}

\begin{figure}[h!]
    \centering
    \includegraphics[width=1\textwidth]{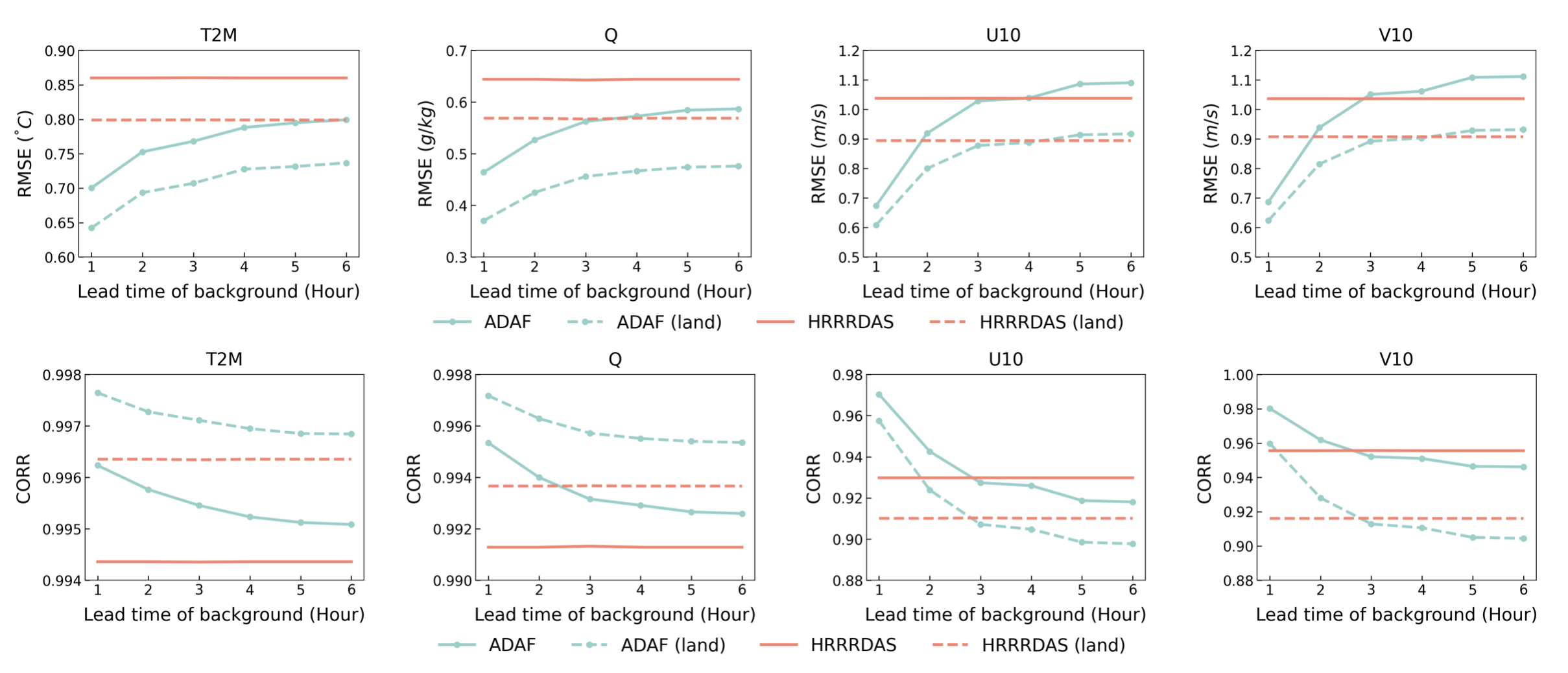}
    \caption{Relationship between analysis errors and the background accuracy. The background accuracy decrease with the lead time increase. The errors, represented as the domain-averaged RMSE and CORR, are calculated by comparison with RTMA. The ADAF analysis shows superior performance to the HRRRDAS analysis by achieving lower RMSE and higher CORR values for most lead times. 
    Errors in ADAF analysis rise when using forecast backgrounds with longer lead time, emphasizing the need for accurate backgrounds for successful DA. ADAF is capable of producing high-quality analysis even when utilizing backgrounds with low accuracy, thus enhancing the timeliness of the DA process.}
    \label{fig:sensitive_to_lead_time}
\end{figure}

To assess ADAF's sensitivity to sparse surface observations, we tested it with different observation sparsity levels. The sparsity of observation is represented as the observation grid ratio, as defined in Equation~\ref{eq:obs_grid_ratio}. 
The ratio varied between 0.5\% and 2.2\% in our study. NaNs at grid points without observations were substituted with zeros.

\begin{linenomath}
\begin{equation}
    \text{Observation grid ratio} = \frac{\text{total number of observation grid points}}{\text{total number of grid points (= 512 $\times$ 1280)}}
\label{eq:obs_grid_ratio}
\end{equation}
\end{linenomath}

Figure~\ref{fig:sensitive_to_obs_sparsity} illustrates the relationship between observation sparsity and errors in ADAF analysis. 
The errors are quantified by the domain-averaged RMSE and CORR, compared to RTMA.
The ADAF analysis (depicted with blue lines) generally shows lower RMSE and higher CORR compared to the HRRRDAS analysis (depicted with red lines) at various observation sparsity levels.
The HRRRDAS errors are constant because the sparsity of observations in the HRRRDAS analysis has not changed. 
The results show that the ADAF analysis achieves greater accuracy than the HRRRDAS analysis, which uses EnKF, even when observations are highly sparse (with an observation grid ratio of 0.5\%).
The errors in the ADAF analysis on land are smaller than those in the HRRRDAS analysis, mainly because that surface observations on land provide valuable information to the land state reconstruction.
As surface observations become sparser, the error correspondingly increases, suggesting that a higher density of observations can enhance the accuracy of the analysis. 
Furthermore, ADAF's computational cost remains unchanged even as the number of observations grows, because the data are pre-processed into a tensor for efficient GPU processing.
In summary, ADAF is robust to generating high-quality analysis even when observations are extremely sparse, indicating that the proposed ADAF method can transfer available observations into unobserved regions.

\subsubsection{Sensitive to Background Accuracy}
\label{sec:sensitive_to_background_acc}

To assess the impact of background accuracy on ADAF, we performed a sensitivity analysis using backgrounds derived from different lead times. 
The accuracy of the background declines as the lead time increases, as the quality of weather forecasts deteriorates with a longer lead time due to the chaotic nature of atmospheric processes~\cite{Ramage1975Prognosis, Chantry2019Scale}. 
We train a model for each lead time. 
Figure~\ref{fig:sensitive_to_lead_time} shows the relationship between lead time and analysis errors (represented as RMSE and CORR), which are compared to RTMA. 
The results indicate that ADAF analysis outperforms HRRRDAS analysis, generated via EnKF, with reduced RMSE and increased CORR for most lead times. Although HRRRDAS errors remain constant with a 1-hour forecast as background, ADAF demonstrates superior accuracy over HRRRDAS under the same conditions.
ADAF analysis errors increase when background with longer lead times is used, highlighting the importance of background accuracy for analysis generation. 
Notably, these errors rise substantially for longer lead time backgrounds in U10 and V10, suggesting wind field reconstruction heavily depends on background accuracy.
In summary, ADAF is capable of producing high-quality analysis even when utilizing backgrounds with low accuracy, thus enhancing the timeliness of the DA process.

\subsubsection{Cases Visualization}

\begin{figure}[h!]
    \centering
    \includegraphics[width=1\textwidth]{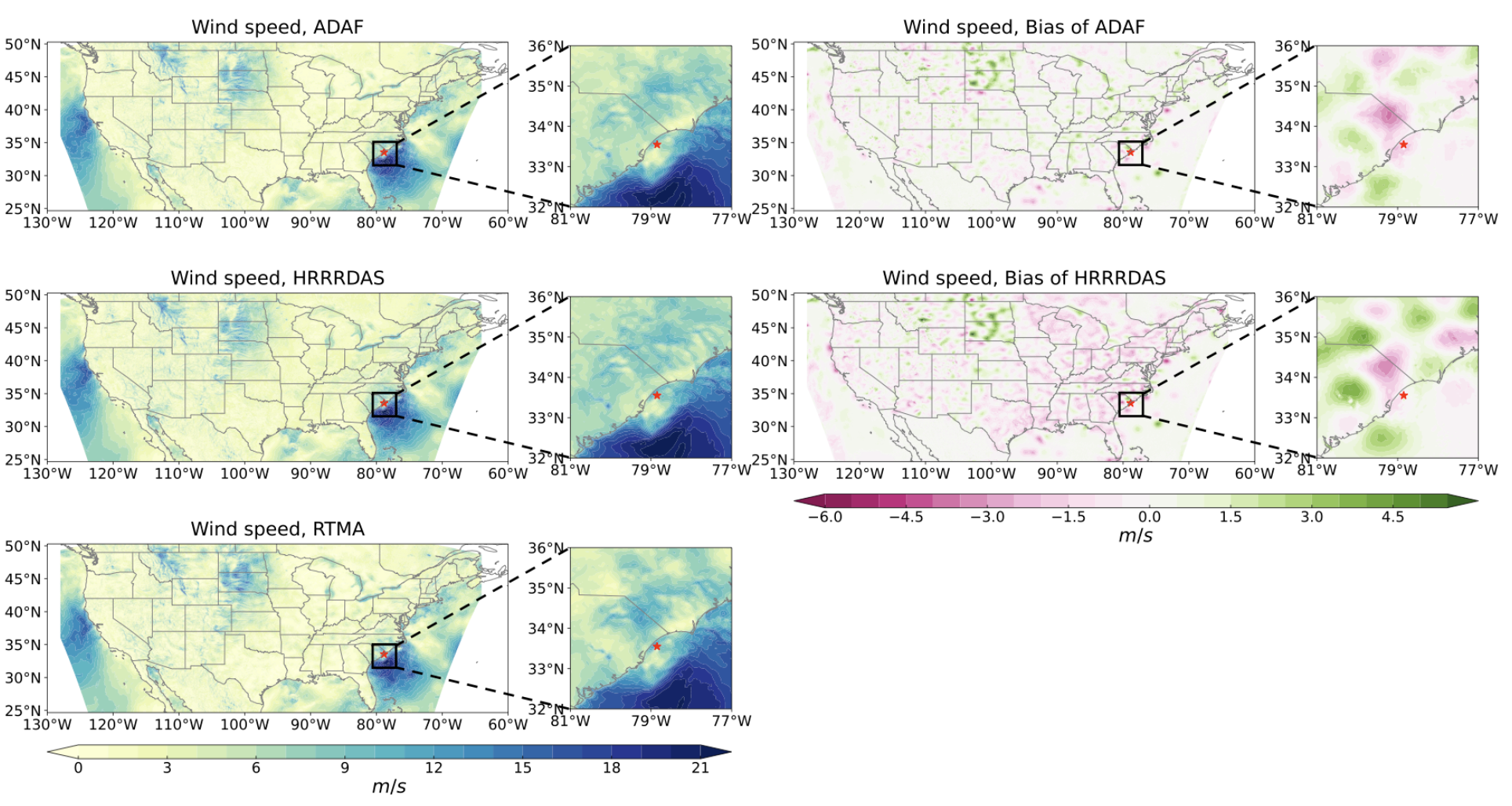}
    \caption{Wind speed maps for ADAF analysis, HRRRDAS analysis, RTMA, and the biases of ADAF and HRRRDAS analysis when comparing with RTMA.
    The analysis time is 06:00 UTC 31 Aug 2023. A 4 $^\circ$ $\times$ 4$^\circ$ region around Typhoon IDALIA (33.55$^\circ N$, 78.85$^\circ W$) has been amplified. The bias of ADAF analysis in this region is significantly smaller than that of HRRRDAS analysis, with the maximum MAE reducing from 5.15 to 4.41 $m/s$ and the region-averaged MAE reducing from 1.17 to 0.84 $m/s$.
    The result show that ADAF can reconstruct wind field of tropical cyclone by assimilating satellite imagery.}
    \label{fig:Typhon_case_IDALIA-2023-08-31_06}
\end{figure}

To assess ADAF's capacity to reconstruct atmospheric state during extreme wind events, we selected several tropical cyclone cases from the International Best Track Archive for Climate Stewardship (IBTrACS)~\cite{Gahtan_2024_IBTrACS, Kenneth_2010_IBTrACS}, which provides the most comprehensive global collection of tropical cyclones. 
Figure~\ref{fig:Typhon_case_IDALIA-2023-08-31_06} illustrates the reconstruction of wind speed in a 4 $^\circ$ $\times$ 4$^\circ$ region centered on the tropical cyclone IDALIA (33.55$^\circ N$, 78.85$^\circ W$) at 06:00 UTC on August 31, 2023. 
The results show that the bias of ADAF analysis in this region is significantly smaller than that of HRRRDAS analysis, with the maximum MAE reducing from approximately 5.15$m/s$ to 4.41$m/s$ and the region-averaged MAE reducing from 1.17$m/s$ to 0.84$m/s$. 
The results underscore ADAF's capability to reconstruct atmospheric states during extreme weather events by assimilating GOES-16 satellite imagery.
More cases are shown in Figures~\ref{SI-fig:Typhon_case_IDALIA-2023-08-30_18} and \ref{SI-fig:Typhon_case_OPHELIA-2023-09-23_18}.

\FloatBarrier

\begin{figure}[h!]
    \centering
    \includegraphics[width=1\textwidth]{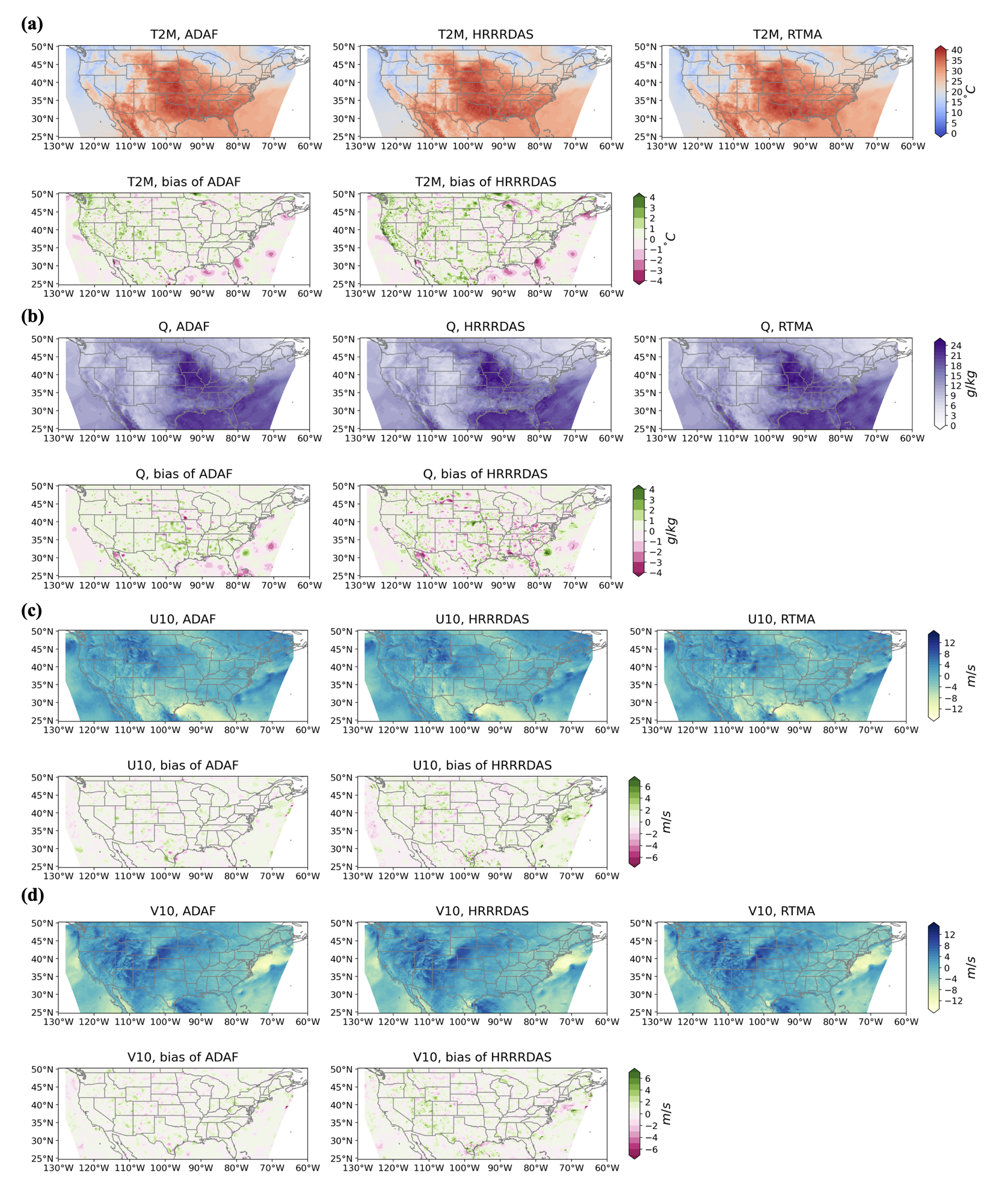}
    \caption{Maps of the ADAF, HRRRDAS and RTMA analysis, the biases of ADAF and HRRRDAS analysis against RTMA. Four near-surface variables were visualized, including (a) T2M, (b) Q, (c) U10, and (d) V10. The analysis time is 18:00 UTC on August 22, 2023. The results show that the bias of ADAF analysis is significantly smaller than that HRRRDAS analysis, with the region-averaged MAE decreasing from 0.65 to 0.52 $^\circ C$ for T2M, from 0.63 to 0.47 $g/kg$ for Q, from 0.73 to 0.54 $m/s$ for U10, and from 0.72 to 0.55 $m/s$ for V10. The case indicates that ADAF is highly effective in reconstructing the atmospheric states of near-surface variables by assimilating multi-source observations.}
    \label{fig:case_2023-08-22_18}
\end{figure}

Moreover, we investigated some cases for four near-surface variables: T2M, Q, U10 and V10. 
Figure~\ref{fig:case_2023-08-22_18} presents the analysis generated by ADAF and HRRRDAS, along with RTMA.
The spatial bias distribution for ADAF and HRRRDAS analysis is also visualized. The analysis time is 18:00 UTC on August 22, 2023.
The ADAF analysis shows a high level of consistency with RTMA spatial patterns. 
And the bias of ADAF analysis is significantly smaller than that of HRRRDAS analysis, with the average MAE reducing from 0.65 to 0.52 $^\circ C$ for T2M, from 0.63 to 0.47 $g/kg$ for Q, from 0.73 to 0.54 $m/s$ for U10, and from 0.72 to 0.55 $m/s$ for V10, respectively.
In summary, the proposed ADAF shows great efficacy in reconstructing atmospheric states by successfully assimilating multi-source observations. More cases are shown in Figures~\ref{SI-fig:case_2023-09-23_12} and \ref{SI-fig:case_2023-07-12_12}.

\FloatBarrier

\section{Conclusion and Discussion}
\label{sec:conclusion_and_discussion}

Data assimilation (DA) is essential in weather forecasting, integrating observational data into numerical weather prediction (NWP) models to enhance forecast accuracy. Traditional DA methods, such as variational methods (3D-Var and 4D-Var), and ensemble Kalman filters (EnKF) have significantly improved forecast reliability. These methods effectively combine model simulations with observations, correcting model states, and reducing uncertainties. However, they often face challenges in balancing cost-effectiveness and accuracy due to complex linear algebra computations and the high dimensionality of the model, especially in nonlinear systems. Moreover, processing vast amounts of real-time data requires substantial computational resources.
The increasing volume and variety of observational data, along with the push for higher-resolution models, have intensified these challenges, prompting the exploration of more efficient DA approaches.

In this study, we introduce an AI-based data assimilation framework (ADAF), which can generate high-quality analysis on a kilometer scale. 
This study is the first pioneering work that uses real-world observations from multiple sources and varied locations to verify the AI method's efficacy in DA, including sparse surface weather observations and satellite imagery.
We applied ADAF for four variables near the surface in the CONUS. 
The results show that ADAF outperforms HRRRDAS in depicting near-surface atmospheric conditions, aligning more closely with actual observations. Accuracy improvements are from 16\% to 33\% over RTMA and 5.7\% to 7.7\% over observations.
Sensitivity tests demonstrate that ADAF can deliver reliable analysis even with low-quality backgrounds and extremely sparse surface data. ADAF can reconstruct tropical cyclone wind fields and integrate a large number of observations within a three-hour window at low computational cost, taking about two seconds on an AMD MI200 GPU. ADAF has shown efficiency and effectiveness in real-world data assimilation, underscoring its promise for operational weather prediction.

Additional research is required to verify the effectiveness of ADAF in weather forecasting systems. The subsequent step is to evaluate the forecast performance using ADAF analysis as initial conditions in numerical or AI-based weather models.
In addition, incorporating uncertainty quantification into ADAF is crucial for reducing errors and enabling ensemble forecasting.
Assimilating diverse observations such as radar and satellite data enhances the precision of estimation fields, especially in severe weather situations. Thus, by using multisource observations, similar to those in NWP's DA system, ADAF can produce detailed atmospheric analysis fields.
In this study, ADAF was initially applied to CONUS due to data limitations. However, its low sensitivity to input sparsity and background field quality suggests that it could adapt to other regions and globally.
Moreover, ADAF's capability to incorporate surface observations suggests that by integrating vertical atmospheric observations—such as satellite profiles and GPS Radio Occultation, which provide vertical information on temperature and humidity—it can produce a 3D atmospheric analysis.
The overall success of the approach introduced in this study is expected to encourage additional efforts in an operational weather forecasting setting that leverage AI techniques.

\section*{Data Availability Statement}

Some of the pre-processed data and the weights of the trained model can be accessed~\cite{xiang_2024_14020879}. 
The code for executing ADAF is openly accessible at \url{https://github.com/xiangyanfei212/ADAF.git}.

\acknowledgments
The authors gratefully acknowledge NOAA for publicly accessible RTMA analysis data and GOES satellite data. The authors appreciate Synoptic Data PBC (linked to \url{https://synopticdata.com/}) for aggregating real-world weather station observations, which were essential in developing an AI-based framework for data assimilation. This work was supported by the National Natural Science Foundation of China (42125503).

\appendix
\section{High Resolution Rapid Refresh Data Assimilation System (HRRRDAS)}
\label{SI-sec:HRRRDAS}

\begin{figure}[h!]
    \centering
    \includegraphics[width=0.9\textwidth]{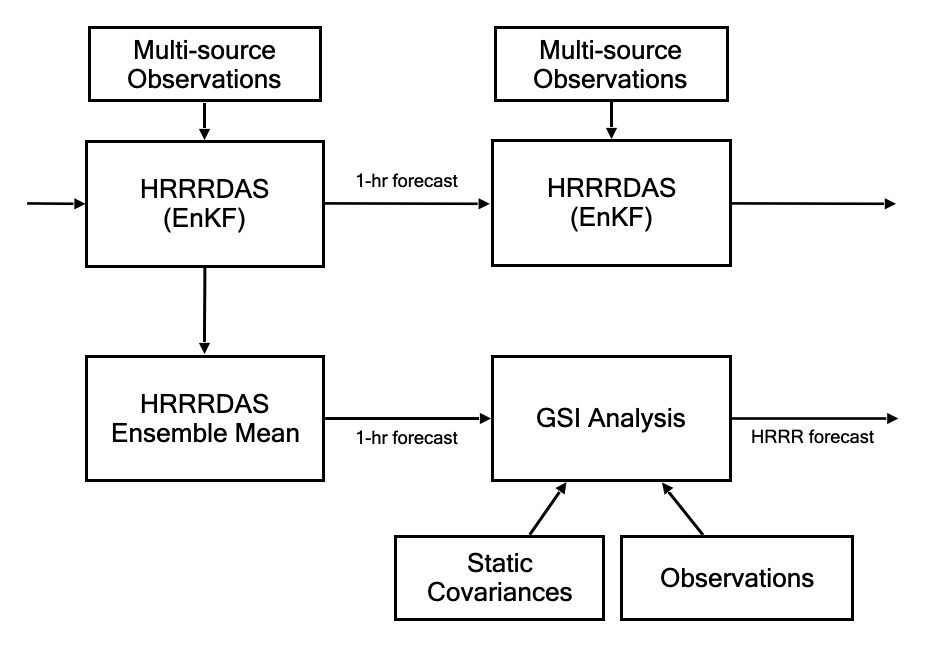}
    \caption{Pipeline for hourly-cycled HRRRDAS~\cite{Dowell_2022_HRRR}. Blue, green, and red indicate deterministic, ensemble, and other information (observations and static covariances), respectively.}
    \label{SI-fig:HRRRDAS_pipeline}
\end{figure}
\FloatBarrier

HRRRv4 exhibits a significant improvement in DA compared to previous versions through the use of the HRRR 3-km data assimilation system (HRRRDAS). HRRRDAS provides the initial condition for HRRRv4~\cite{Dowell_2022_HRRR}. It employs the ensemble Kalman filter (EnKF)~\cite{Houtekamer_Review_2016}, specifically the ensemble square-root filter (EnSRF)~\cite{EnsembleSquareRootFilters}, to integrate multi-source observations, including surface measurements, satellite imagery from GOES-16, and radar reflectivity data from NOAA's Multi-Radar Multi-Sensor (MRMS) project. 
The HRRR initialization comprises an initial guess, a one-hour 'forecast', and the Gridpoint Statistical Interpolation (GSI) hybird data assimilation system. 
For HRRRv4 CONUS initialization, the HRRRDAS ensemble mean provides the first guess, and the HRRRDAS 1-h forecasts provide the background error covariances for the GSI hybrid data assimilation system, as depicted in Figure~\ref{SI-fig:HRRRDAS_pipeline}.
HRRRDAS follows an hourly update cycle, which repeats the DA process each hour.

\section{Near Surface Measurements}

\begin{figure}[h!]
    \centering
    \includegraphics[width=0.4\textwidth]{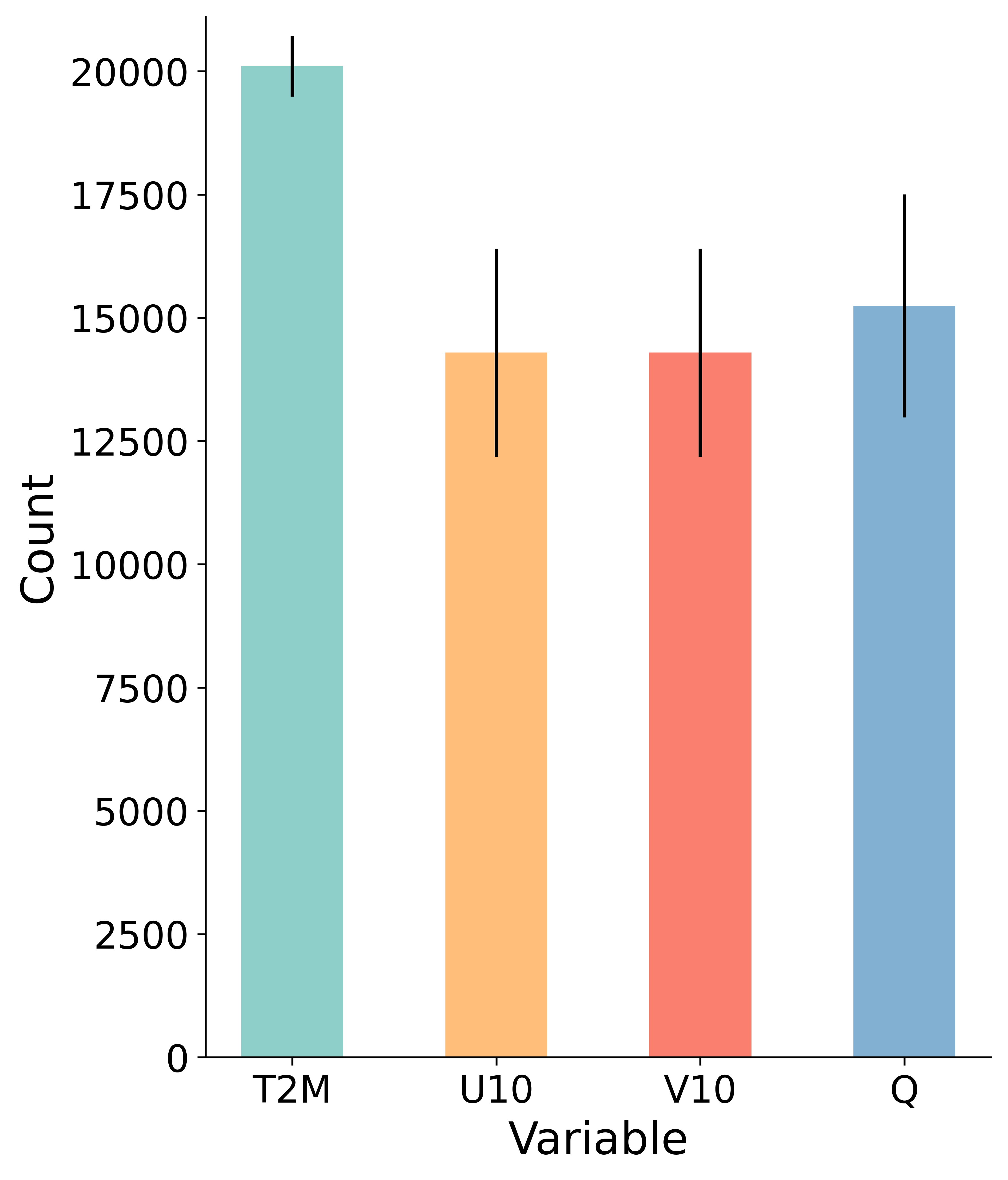}
    \caption{Hourly observation counts for four near-surface atmospheric variables: T2M, U10, V10, and Q.}
    \label{SI-fig:number_of_meansurements}
\end{figure}

\begin{figure}[h!]
    \centering
    \includegraphics[width=0.9\textwidth]{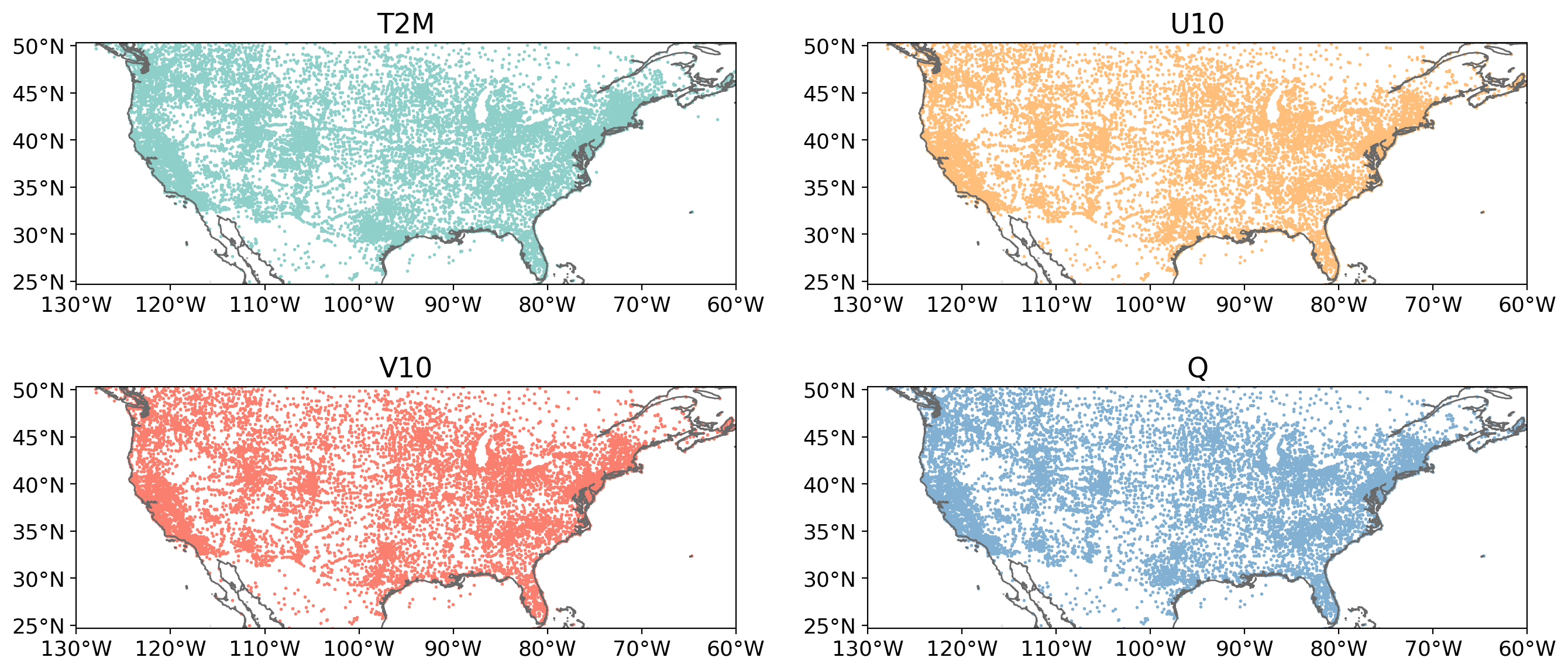}
    \caption{The spatial distribution of surface weather observations for four near-surface atmospheric variables: T2M, U10, V10, and Q.}
    \label{SI-fig:spatial_dist_of_meansurements}
\end{figure}

Surface weather observations used in this study are sourced from WeatherReal-Synoptic~\cite{jin2024weatherreal}, which gives quality control to station measurements collected by Synoptic Data.
For more details, please visit the Synoptic Data official website~\cite{synoptic}.
Figure~\ref{SI-fig:number_of_meansurements} depicts the average hourly observations, while Figure~\ref{SI-fig:spatial_dist_of_meansurements} shows the spatial distribution.

\FloatBarrier

\FloatBarrier

\section{Tropical cyclone wind speed field}

\begin{figure}[h!]
    \centering
    \includegraphics[width=0.95\textwidth]{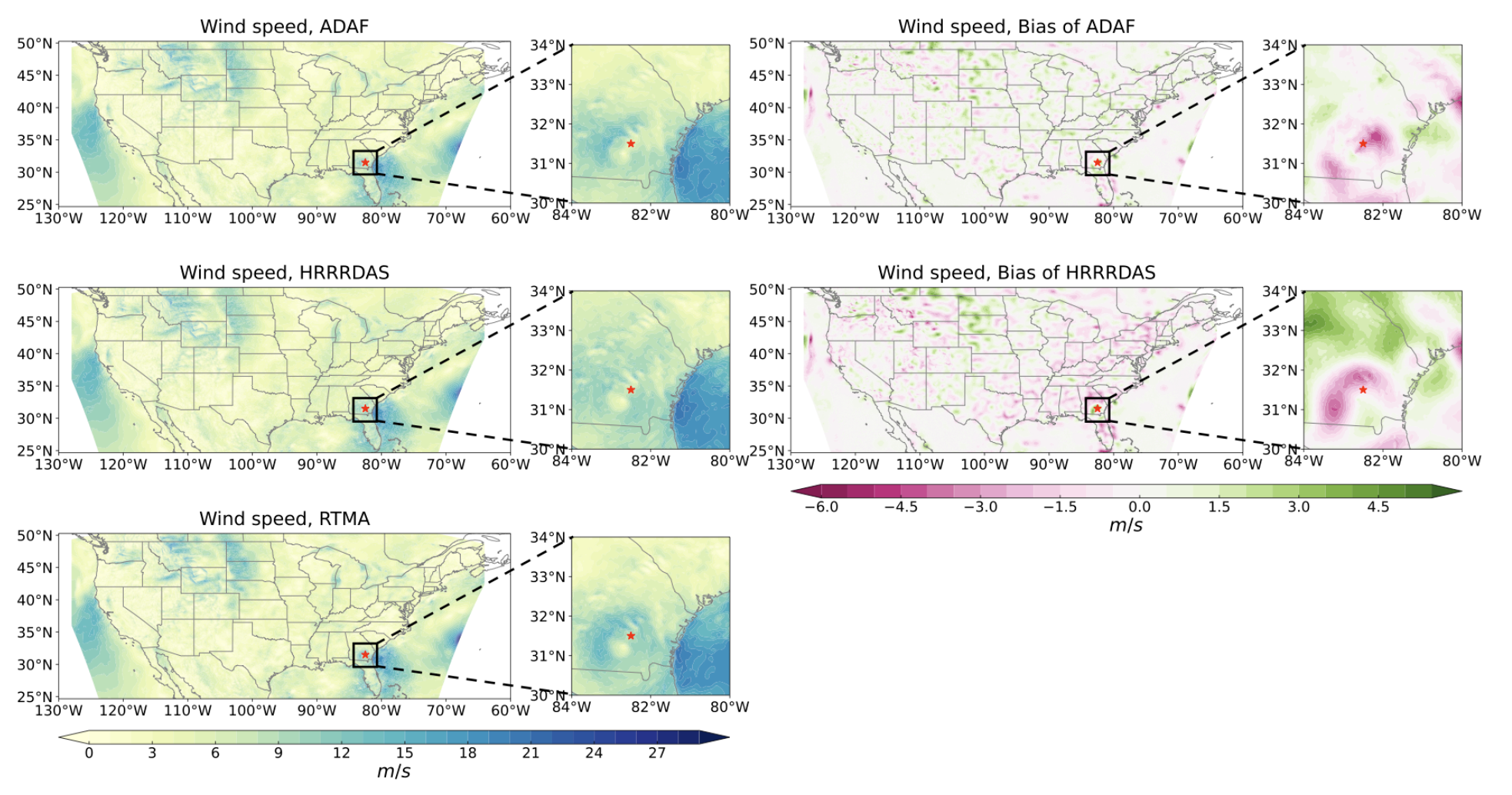}
    \caption{Wind speed maps for ADAF analysis, HRRRDAS analysis, RTMA, and the biases of ADAF and HRRRDAS analysis when comparing with RTMA.
    The analysis time is 18:00 UTC 30 Aug 2023. A 4 $^\circ$ $\times$ 4$^\circ$ region around Typhoon IDALIA (31.5$^\circ N$, 82.5$^\circ W$) has been amplified. The bias of ADAF analysis in this region is significantly smaller than that of HRRRDAS analysis, with the maximum MAE reducing from 9.87 to 7.09 $m/s$ and the region-averaged MAE reducing from 1.35 to 1.08 $m/s$.}
    \label{SI-fig:Typhon_case_IDALIA-2023-08-30_18}
\end{figure}

\begin{figure}[h!]
    \centering
    \includegraphics[width=0.95\textwidth]{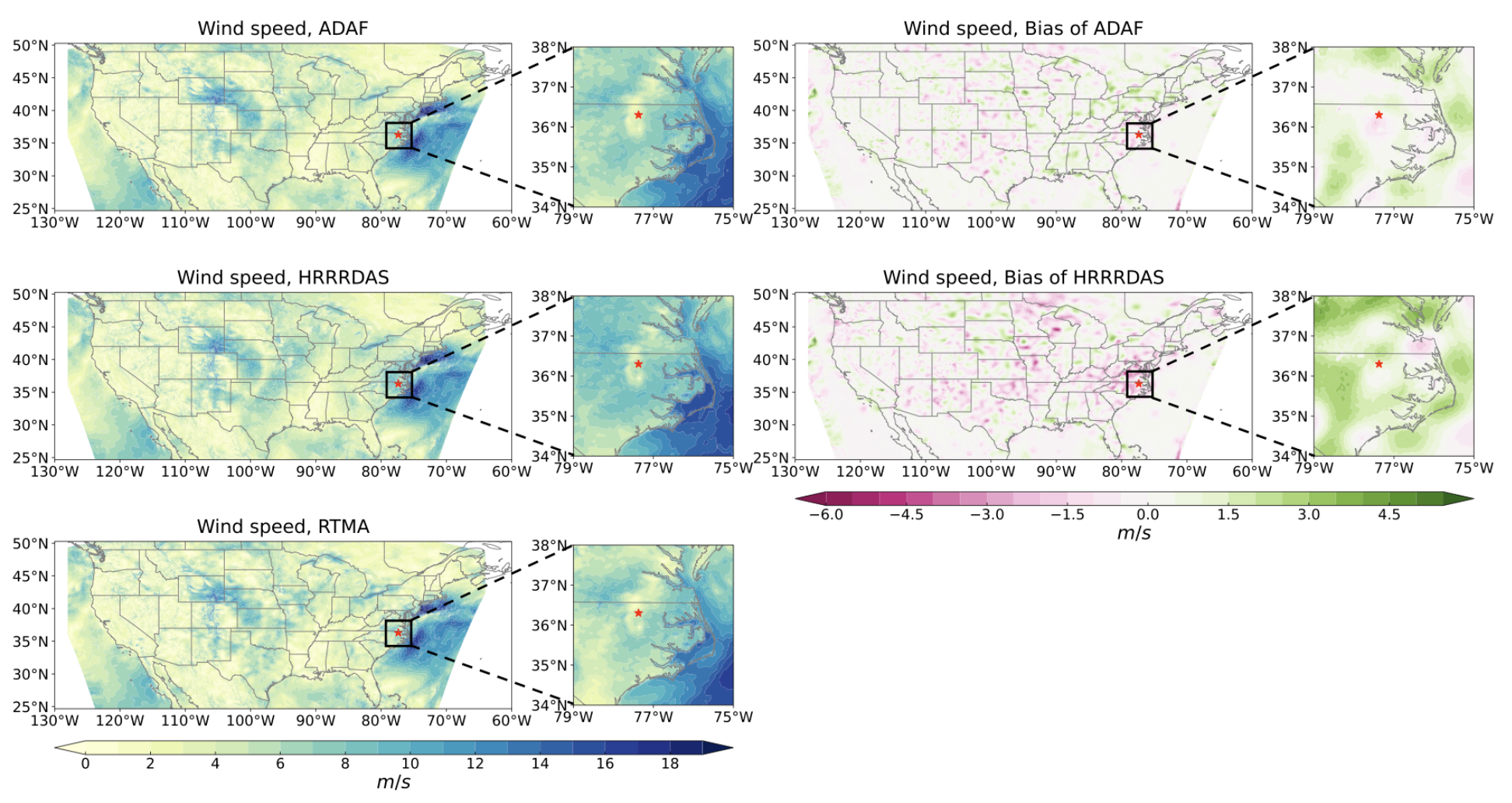}
    \caption{Wind speed maps for ADAF analysis, HRRRDAS analysis, RTMA, and the biases of ADAF and HRRRDAS analysis when comparing with RTMA.
    The analysis time is 18:00 UTC 23 Sep 2023. A 4 $^\circ$ $\times$ 4$^\circ$ region around Typhoon OPHELIA (36.3$^\circ N$, 77.4$^\circ W$) has been amplified. The bias of ADAF analysis in this region is significantly smaller than that of HRRRDAS analysis, with the maximum MAE reducing from 8.72 to 6.63 $m/s$ and the region-averaged MAE reducing from 1.62 to 0.93 $m/s$.}
    \label{SI-fig:Typhon_case_OPHELIA-2023-09-23_18}
\end{figure}

To assess ADAF's capacity to reconstruct atmospheric state during extreme wind events, we selected several tropical cyclone cases from the International Best Track Archive for Climate Stewardship (IBTrACS)~\cite{Gahtan_2024_IBTrACS, Kenneth_2010_IBTrACS}, which provides the most comprehensive global collection of tropical cyclones. 
Figures~\ref{SI-fig:Typhon_case_IDALIA-2023-08-30_18} and ~\ref{SI-fig:Typhon_case_OPHELIA-2023-09-23_18} illustrates the reconstruction of wind speed in a 4 $^\circ$ $\times$ 4$^\circ$ region centered on the tropical cyclone IDALIA and OPHELIA.

\FloatBarrier

\FloatBarrier

\section{Visualization of the near-surface variables} 

\begin{figure}[h!]
    \centering
    \includegraphics[width=1\textwidth]{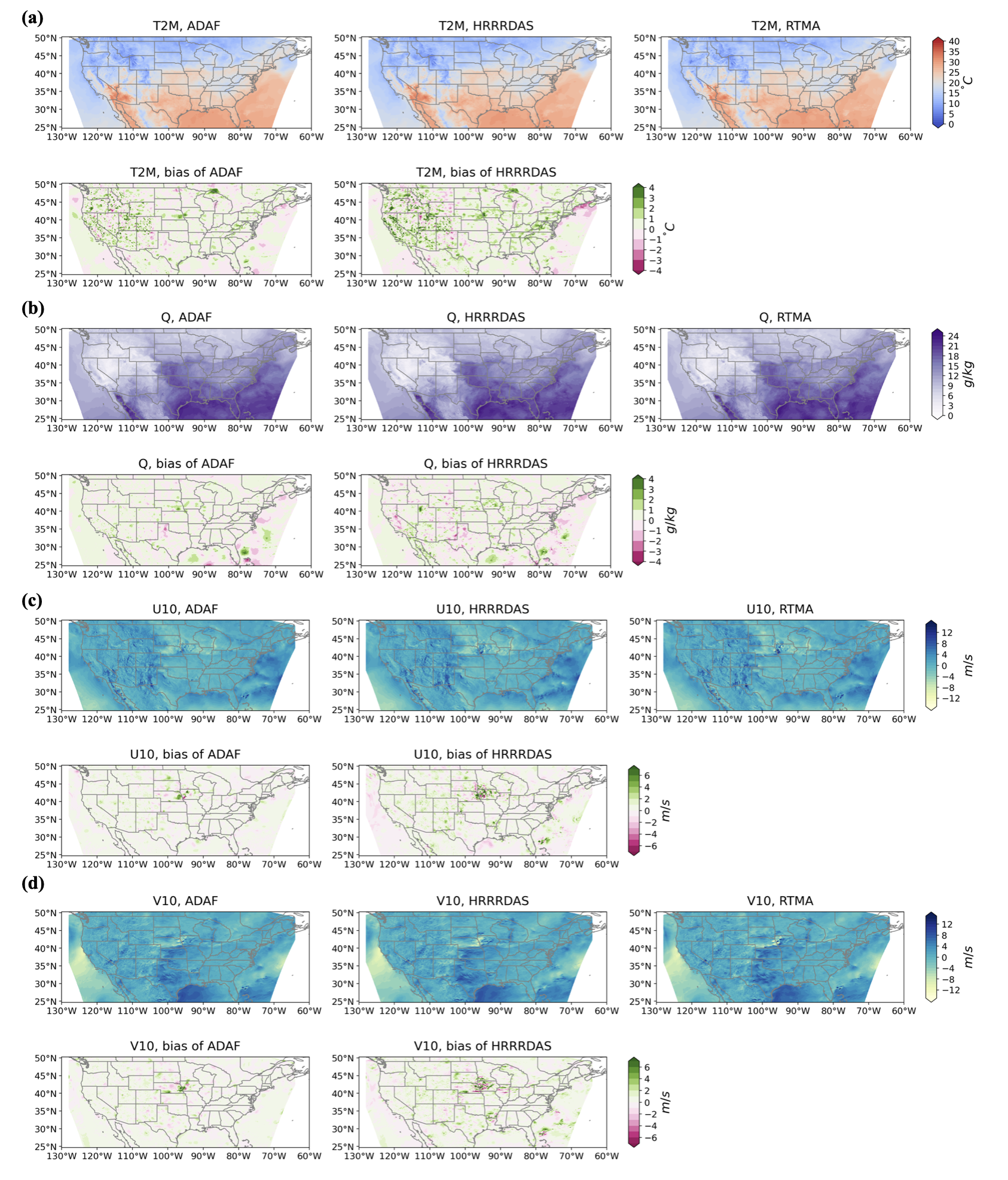}
    \caption{Maps of the ADAF, HRRRDAS and RTMA analysis, the biases of ADAF and HRRRDAS analysis against RTMA. Four near-surface variables were visualized, including (a) T2M, (b) Q, (c) U10, and (d) V10. The analysis time is 12:00 UTC on July 12, 2023. The results show that the bias between ADAF analysis and RTMA is significantly smaller than the bias between HRRRDAS analysis and RTMA, with the region-averaged MAE reducing from 0.60 to 0.49 $^\circ C$ for T2M, from 0.47 to 0.36 $g/kg$ for Q, from 0.68 to 0.42 $m/s$ for U10, and from 0.70 to 0.43 $m/s$ for V10.}
    \label{SI-fig:case_2023-07-12_12}
\end{figure}

\begin{figure}[h!]
    \centering
    \includegraphics[width=1\textwidth]{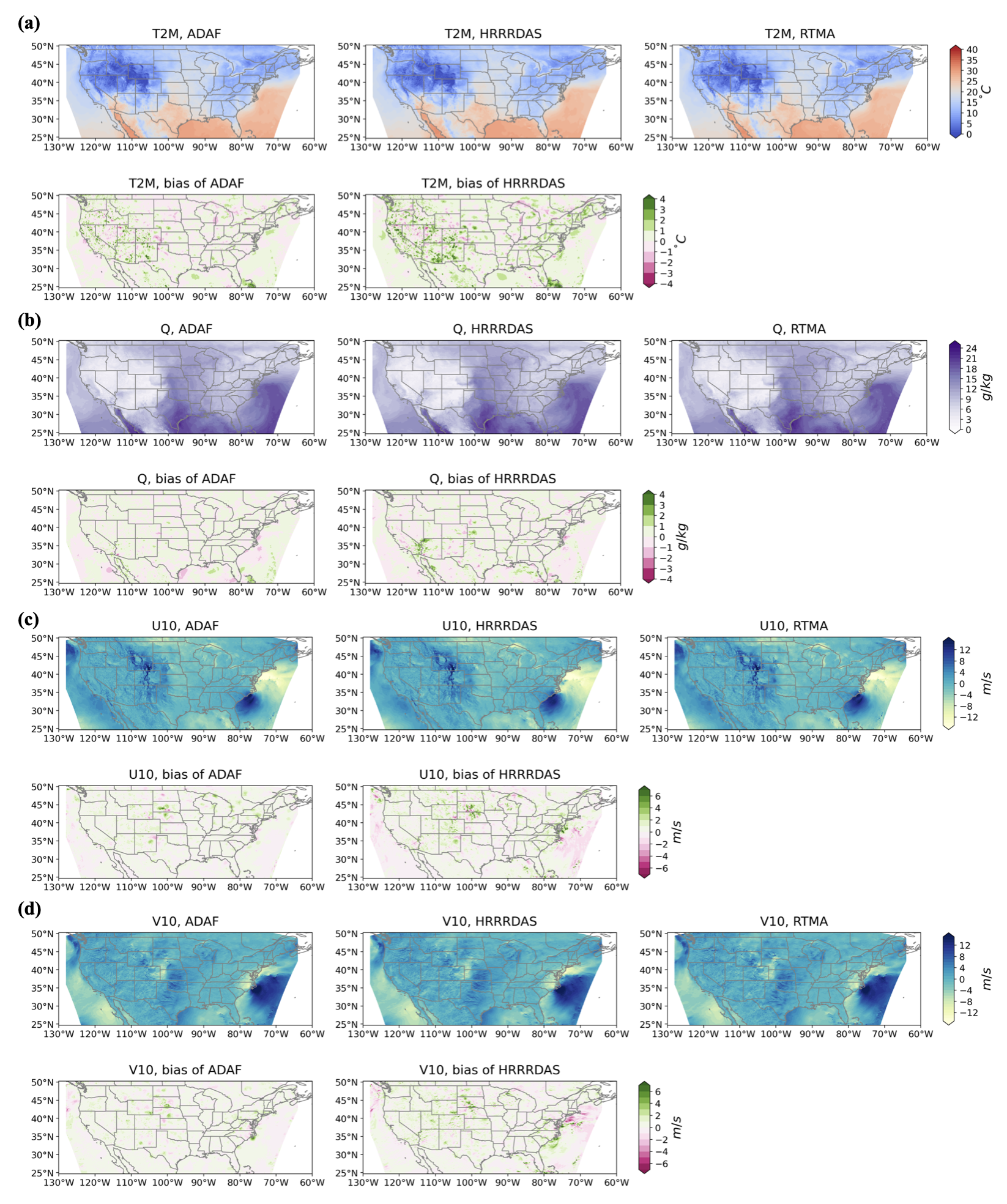}
    \caption{Maps of the ADAF, HRRRDAS and RTMA analysis, the biases of ADAF and HRRRDAS analysis against RTMA. Four near-surface variables were visualized, including (a) T2M, (b) Q, (c) U10, and (d) V10. The analysis time is 12:00 UTC on September 23, 2023. The results show that the bias between ADAF analysis and RTMA is significantly smaller than the bias between HRRRDAS analysis and RTMA, with the region-averaged MAE reducing from 0.55 to 0.44 $^\circ C$ for T2M, from 0.36 to 0.27 $g/kg$ for Q, from 0.72 to 0.45 $m/s$ for U10, and from 0.70 to 0.43 $m/s$ for V10.}
    \label{SI-fig:case_2023-09-23_12}
\end{figure}

Figures~\ref{SI-fig:case_2023-07-12_12} and ~\ref{SI-fig:case_2023-09-23_12} display maps of four near-surface variables from ADAF, HRRRDAS, and RTMA analysis, along with the biases of ADAF and HRRRDAS against RTMA. Four near-surface variables were visualized, including T2M, Q, U10, and V10.

\FloatBarrier

\FloatBarrier

\bibliography{references}


\end{document}